\documentclass[journal]{IEEEtran}
\usepackage{amsfonts}
\IEEEoverridecommandlockouts

\ifCLASSINFOpdf
\else
\fi
\usepackage{epstopdf}
\usepackage{multirow}
\usepackage{cite}
\usepackage{stfloats}
\usepackage{color}
\usepackage{epsfig}
\usepackage{graphicx}
\usepackage{psfig}
\usepackage{epsf}
\usepackage{slashbox}
\usepackage{float}
\usepackage{amssymb }
\usepackage[cmex10]{amsmath}
\begin{document}

\title{System Outage Performance for Three-Step Two-Way Energy Harvesting DF Relaying}
\author{Liqin Shi, Yinghui Ye, Rose Qingyang Hu,~\IEEEmembership{Senior Member,~IEEE,} and Hailin Zhang,~\IEEEmembership{Member,~IEEE}
\thanks{Copyright (c) 2015 IEEE. Personal use of this material is permitted. However, permission to use this material for any other purposes must be obtained from the IEEE by sending a request to pubs-permissions@ieee.org.}
\thanks{The research reported in this article was supported by the scholarship from China Scholarship Council, the National Natural Science Funding of China under grant 61671347, the Science and Technology Innovation Team of Shaanxi Province for Broadband Wireless and Application under grant 2017KCT-30-02, and the 111 Project of China under grant B08038. The work of Prof. R. Q. Hu was supported by the US National Science
Foundation under the Grants NeTS-1423348 and EARS-1547312.}
\thanks{Liqin Shi, Yinghui Ye and Hailin Zhang  are with the State Key Laboratory of Integrated Service Networks, Xidian University, Xi'an 710071, China (e-mails: liqinshi@hotmail.com, connectyyh@126.com, hlzhang@xidian.edu.cn).}
\thanks{Rose Qingyang Hu  is with the Department of Electrical and Computer Engineering, Utah State University, USA (e-mail: rose.hu@usu.edu).}
}
\markboth{IEEE Transactions on Vehicular Technology, No. XX, MONTH YY, YEAR 2019}
{Shi\MakeLowercase{\textit{et al.}}: Outage Performance Optimization for SWIPT Enabled Three-Step Two-Way DF Relaying}
\maketitle

\begin{abstract}
Wireless energy harvesting (WEH) has been recognized as a promising technique to prolong the lifetime of energy constrained relay nodes in wireless sensor networks. Its application and related performance study in three-step two-way decode-and-forward (DF) relay networks are of high interest but still lack sufficient study. In this paper
 we propose a dynamic power-splitting (PS) scheme to minimize the system outage probability in a three-step two-way energy harvesting  DF relay network
and derive an analytical expression for the system outage probability with respect to the optimal dynamic PS ratios.
In order to further improve the system outage performance, we propose an improved dynamic scheme where both the PS ratios and the power allocation ratio at the relay are dynamically adjusted according to instantaneous channel gains. The corresponding system performance with the improved dynamic scheme is also investigated. Simulation results show that our proposed schemes outperform the existing scheme in terms of the system outage performance and the improved dynamic scheme is superior to the dynamic PS scheme.
\end{abstract}
\IEEEpeerreviewmaketitle
\section{Introduction}
\IEEEPARstart{A}{s} the era of Internet of Things (IoT) approaches, an explosive growth of IoT devices, such as  low-power wireless devices in wireless sensor networks,  will be connected into the network to share and forward information, bringing intelligence and convenience to our life \cite{7113786}.
One of the key challenges to realize IoT is how to power up the massive  number of devices while maintaining the required
quality of service \cite{8187650,7744827}. Radio frequency (RF) based wireless energy harvesting (WEH) has been recognized as an effective solution  to address this challenge.  By exploiting the dual use of RF signals,
WEH could be integrated with wireless communications to yield a new technology, namely simultaneous wireless information and power transfer (SWIPT), where RF signals are either switched in the time domain or split in the power domain to facilitate energy harvesting and information transmission through a time-switching (TS) scheme or power-splitting (PS) scheme \cite{6957150}.

On the other hand, wireless relaying has been widely employed in current and emerging wireless systems for efficient information transmission by cutting down multipath fading and shadowing and increasing the diversity order \cite{5534586}. For example, wireless relaying can  be utilized extensively  in machine-to-machine networks,  where low power  IoT devices exchange their data through an immediate relay \cite{6664485}. In conventional relay networks, a relay provides free services and  costs its extra energy, which may prevent energy-constrained nodes from engaging.
Thus, the aforementioned two communication concepts, SWIPT and wireless relaying, can be combined to  motivate  relays to assist data exchange \cite{8315209,7120018,ye11}.
In particular, SWIPT can be  built upon basic one-way relaying \cite{ye11,8012410,AA13,8337780} or two-way relaying \cite{7565756,7831382,7032100, T23,7971948,7997272,7565032,7858680,EL,8287997,8361446,2017CL,8377371}.
Two-way relaying can be performed in two steps or three steps compared with four steps as required in one-way relaying. Thus, SWIPT enabled two-way relaying can be  more spectrally efficient than SWIPT enabled one-way relaying.  In this regard, SWIPT enabled two-way relaying has received increasing attention and has been investigated extensively \cite{7565756, 7831382,7032100, T23,7971948,7997272,7565032,7858680,EL,8287997,8361446,2017CL,8377371}.

In \cite{7565756}, the authors studied the outage probability for  three wireless
power transfer schemes in TS based two-step amplify-and-forward (AF) two-way relay networks (TWRNs).   In another study \cite{7831382,7032100} the outage behavior and finite signal-to-noise ratio (SNR) diversity multiplexing trade-off  were analyzed. In contrast, by considering a decode-and-forward (DF) protocol in SWIPT enabled two-step  TWRNs, the authors of  \cite{T23} proposed a resource allocation scheme to minimize the system outage probability by  jointly optimizing the time allocation ratio and the PS/TS ratio. A comprehensive comparison between SWIPT enabled two-step DF TWRNs and SWIPT enabled two-step AF TWRNs was presented in \cite{7997272,7565032}.

Recall that the low hardware complexity is very critical to energy-constrained relay networks and the  circuit design of three-step two-way relaying is simpler than that of two-step two-way relaying. Several studies \cite{EL,8287997,8361446,2017CL,8377371,shi} on the SWIPT enabled three-step two-way relaying have been hitherto reported.
The authors in \cite{EL} investigated a static equal PS scheme to maximize the system outage capacity for AF based three-step TWRNs, where the PS ratio is determined by the statistical channel state information (CSI).
Since the outage capacity can be improved by adopting a dynamic PS scheme where the PS ratio can be adaptive to the instantaneous CSI instead of the statistical CSI, the dynamic equal PS scheme was further developed \cite{8287997}.
Considering asymmetric instantaneous channel gains between relay and two terminals, the authors in \cite{8361446} proposed a novel dynamic asymmetric PS scheme to minimize the system outage probability for the three-step AF TWRNs, where the PS ratio for each link is designed based on its instantaneous CSI.
Due to the fact that the DF relay is found to be of more practical interest,
the authors in \cite{2017CL, shi} introduced the DF protocol instead of AF protocol into SWIPT enabled three-step TWRNs
and studied the  end-to-end (E2E) outage performance for SWIPT enabled three-step DF TWRNs under the guidance of the static equal and dynamic PS schemes, where the linear  and non-linear EH models are considered, respectively. The outage events at different terminals were considered independently.
However, to the best of our knowledge, there is no open work to investigate \emph{system outage performance} for SWIPT enabled three-step DF TWRNs. It is worth emphasizing that the system outage performance is an important metric that jointly considers the outage events of both E2E links and evaluates the transmission performance of the two E2E links as a whole \cite{7831382,5605920,5738653,6828809}. 

In this paper, we consider a SWIPT enabled three-step  DF TWRN in which both the PS scheme and the \lq\lq harvest-then-forward\rq\rq $\; $strategy are employed.
Please note that, unlike the SWIPT enabled three-step two-way relaying in \cite{EL,8287997,8361446,2017CL,8377371, shi}, our considered network can allocate time resources to relay and terminals in unequal portions for better delivery of information.
Then we investigate the system outage performance for the considered network.
Compared with the study of E2E outage performance in \cite{2017CL, shi}, the analysis on system outage performance is much more challenging
due to the fact that two E2E links are highly correlated.{\footnote{{\color{black}The main differences between our work and the existing work \cite{7565756} are as follows. First, \cite{7565756} considered a TS SWIPT enabled two-way AF relay network while our work focuses on the design of PS SWIPT enabled two-way DF relay networks.
Second, \cite{7565756} focused on the derivations of the E2E/system outage probability under three
wireless power transfer schemes. In our work, we focus on the design of PS scheme and
combining strategy to minimize the system outage probability. The expressions of system outage probability are derived to characterize the performance of the proposed schemes.}}}

The major contributions of this paper are summarized as follows.
\begin{itemize}
\item By exploiting the asymmetric instantaneous channel gains between relay and  two terminals, we propose a dynamic PS scheme to minimize the overall system outage probability, where the PS ratio for each link can be adapted  to its instantaneous CSI. Specifically, the closed-form expressions for the optimal PS ratios are derived. Compared with the static equal PS scheme in \cite{2017CL}, the dynamic PS scheme can provide more flexibility and utilize the instantaneous CSI more effectively.
\item We consider the combining strategy for combining the decoded signals at the relay. In particular, the combining strategy is facilitated by the value of the power allocation ratio at the relay.  Integrating the combining strategy with the dynamic PS scheme, we  develop an improved dynamic scheme to achieve the minimum system outage probability, and derive the optimal solutions in closed forms.
\item  To characterize the performance of the proposed schemes, we derive the analytical expressions of system outage probabilities for the two proposed schemes, respectively. 
    The expressions depict the dependence of the system outage probabilities on parameters such as the transmission power, the power allocation ratio (for the dynamic PS scheme only), the time allocation ratio, the transmission rate, etc, which provides valuable insights in selecting a proper system parameter (e.g., the time allocation ratio).
\item Comparing the improved dynamic scheme, the dynamic PS scheme and the static equal PS scheme, we confirm that the improved dynamic scheme achieves the lowest outage probability and the highest outage capacity, especially for  the case with a larger channel gain difference between relay and two terminals.
\end{itemize}

The remainder of this paper is organized as follows. The system model is provided in Section \uppercase\expandafter{\romannumeral 2}. In Section \uppercase\expandafter{\romannumeral 3}, we propose a dynamic PS scheme to minimize the system outage probability of SWIPT enabled three-step DF TWRNs and  derive the corresponding optimal system outage probability and capacity.
In Section \uppercase\expandafter{\romannumeral 4}, to improve the system outage performance, we further propose an improved dynamic scheme by considering the combining strategy at the relay and the corresponding optimal system outage probability is also derived.
Simulation results are provided in Section \uppercase\expandafter{\romannumeral 5}, followed by conclusions in Section \uppercase\expandafter{\romannumeral 6}.

%

\section{System model}
\begin{figure}
  \centering
  \includegraphics[width=0.4\textwidth]{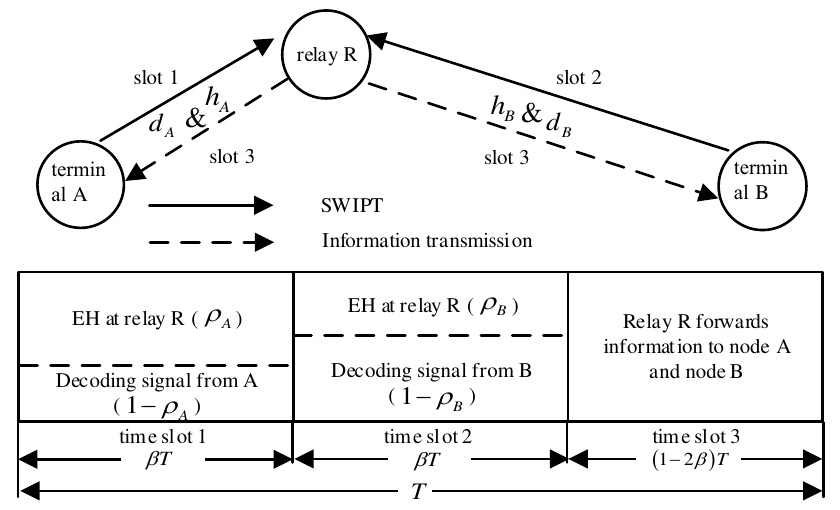}\\
  \caption{System model of the three-step two-way DF relay network.}\label{fig1}
\end{figure}
As shown in Fig. \ref{fig1}, we consider a three-terminal two-way DF relay network,
where terminal \emph{A} communicates with terminal \emph{B} via an energy-constrained relay \emph{R}.
Each terminal has a single antenna and operates in the half-duplex mode.
We assume that no direct link exists between \emph{A} and \emph{B} due to severe path loss and shadowing \cite{EL,8287997,8361446,2017CL}.
Channels are assumed to be reciprocal and quasi-static, and subject to path-loss Rayleigh fading.
Let $h_{A}(h_{B})$ denote the channel coefficient between $A(B)$ and $R$,  and ${d_{ A}}({d_{ B}})$ be the Euclidean distance between $A(B)$ and $R$.
{\color{black}
According to \cite{6951347,8438944}, the path loss model is given by $ \frac{{{G_i}{G_R}{\lambda ^2}}}{{{{\left( {4\pi {d_0}} \right)}^2}}} \times {\left( {\frac{{{d_0}}}{d_i}} \right)^\alpha } \times |h_i|^2$ ($i = A \;\rm{or}\; B$), where $G_i$ is the antenna gain at terminal $i$, $G_R$ is the antenna gain at the relay, $\lambda$ is the carrier wavelength, $d_0$ is the close-in reference distance given from a measurement close to the transmitter and $\alpha $ is the path loss exponent. Further, the path loss model can be rewritten as $\Lambda_i {\left| h_i \right|^2}{d_i^{ - \alpha }}$ ($i = A \;\rm{or}\; B$), where $\Lambda_i=\frac{{{G_i}{G_R}{\lambda ^2}d_0^{\alpha  - 2}}}{{{{\left( {4\pi } \right)}^2}}}$ is a fixed constant for a given scenario.}

{\color{black}Here, we assume that the instantaneous channel coefficients between the two source terminals and the relay are available. Specifically, each source terminal needs to know the instantaneous channel coefficient between the source and the relay. The instantaneous channel coefficient is used to perform successive interference cancellation (SIC) at the source terminal. The relay needs to know the instantaneous channel coefficients between the two source terminals and the relay since the PS ratio for each link is determined by its instantaneous channel coefficient. Note that these instantaneous channel coefficients can be obtained before data transmission in each transmission block. Inspired by \cite{1603719}, we clarify how to obtain these instantaneous channel coefficients as follows. In the investigated system, terminal $A$ broadcasts
a ready-to-send (RTS) message before information transmission. After receiving the RTS message, terminal $B$ replies with a clear-to-send (CTS) message. By overhearing the RTS and CTS messages, relay $R$ can estimate the channel coefficients of both $A$-$R$ and $B$-$R$ channels. Since all the  channels are assumed to be reciprocal, the channel coefficients of both $R$-$A$ and $R$-$B$ channels can be obtained. Terminals $A$ and $B$ are informed of the corresponding channel coefficients through the feedbacks from the relay.}

Let $T$ denote the total transmission block, which is sub-divided into three time slots.
During the first and second time slots with $\beta T$ and $\beta\in(0,0.5)$, $A$ and $B$ transmit their normalized signal $s_{A}$ and $s_{B}$ to $R$ using equal
power\footnote{ Although we make the transmission power at each terminal  the same, the means of $|h_{A}|^2\Lambda_Ad_{A}^{ - \alpha }$ and $|h_{B}|^2\Lambda_Bd_{B}^{ - \alpha }$ are different in general. Accordingly,
average SNRs of all channels can be different, which makes our analysis still
general \cite{1427716}.} $P$, respectively. {\color{black}The received signal from $i$ ($i = A \;\rm{or}\; B$) at $R$ is given by
\begin{align}\label{1}
{y_{ iR}} = {h_{ i}}\sqrt {{P}\Lambda_id_{i}^{-\alpha}} {s_{ i}} + {n_{ iR}},
\end{align}
where
$\mathbb{E}\left\{ {{{\left| s_{i} \right|}^2}} \right\} = 1$ and ${n_{iR}} \sim {\rm{{\cal C}{\cal N}}}\left( {0,\sigma _{iR}^2} \right)$ is the additive white Gaussian noise (AWGN).}

After receiving signal from $i$ ($i = A \;\rm{or}\; B$), $R$ splits it into two parts: $\sqrt{\rho_{i}}{y_{ iR}}$ used for energy harvesting (EH) and $\sqrt{1-\rho_{i}}{y_{ iR}}$ used for information processing. {\color{black}For the energy harvesting, the total harvested energy during first two slots is
\begin{align}\label{2}
E_{\rm{total}}=\beta T\eta P\left( {{{\rho _A}|{h_A}{|^2}\Lambda_Ad_A^{ - \alpha }} + {{\rho _{B}}|{h_{B}}{|^2}\Lambda_Bd_{B}^{ - \alpha }} } \right),
\end{align}
where $\eta$ is the energy conversion efficiency.
For the information processing, the received SNR for decoding $s_{i}$ ($i = A \;\rm{or}\; B$) at the relay is
\begin{align}\label{5}
{\gamma _{iR}} = \frac{{{P}|{h_i}{|^2}\Lambda_i\left( {1 - {\rho _i}} \right)}}{{d_i^\alpha \sigma _{iR}^2}}.
\end{align}}

In the remaining part with $(1-2\beta)T$, $R$ combines the decoded signals $\widetilde{s}_{A}$ and $\widetilde{s}_{B}$ with a power allocation ratio $\theta\in(0,1)$ as $s_{R}=\frac{\theta\widetilde{s}_{A}+(1-\theta)\widetilde{s}_{B}}{\sqrt{\theta^{2}+(1-\theta)^{2}}}$. Note that the value of $\theta$ decides how the relay combines the
decoded signals, $\widetilde{s}_{A}$ and $\widetilde{s}_{B}$.

{\color{black}Then $R$ broadcasts $s_{R}$ to both $A$ and $B$ with the harvested energy $E_{\rm{total}}$ and the received signal at $i$ ($i = A \;\rm{or}\; B$) is given by
\begin{align}\label{6}
{y_{Ri}} &= {h_i}\sqrt {{P_R}\Lambda_id_i^{ - \alpha }} {s_R} + {n_{Ri}},
\end{align}
where ${P_R} = \frac{{{E_{\rm{total}}}}}{{\left( {1 - 2\beta } \right)T}}$ is the transmit power at $R$ and ${n_{Ri}}=\widetilde{n}_{Ri} \sim {\rm{{\cal C}{\cal N}}}\left( {0,\sigma _{Ri}^2} \right)$ is the AWGN at $i$.}
For analytical simplicity, we assume $\sigma _{AR}^2 = \sigma _{BR}^2 = \sigma _{RA}^2 = \sigma _{RB}^2 ={\sigma ^2}$ \cite{2017CL}.

{\color{black}After using SIC at $i$ ($i = A \;\rm{or}\; B$), the end-to-end SNR from $R$ to $i$ is given by
\begin{align}\label{7}
{\gamma _{Ri}} &= {X_i}\left( {{\rho _i}|{h_i}{|^4}\Lambda_id_i^{ - \alpha }{\rm{ + }}{\rho _{\bar i}}|{h_{\bar i}}{|^2}|{h_i}{|^2}\Lambda_{\bar i}d_{\bar i}^{ - \alpha }} \right),
\end{align}
where ${X_i} = \left\{ {\begin{array}{*{20}{c}}
{\frac{{\beta \eta P\Lambda_Ad_A^{ - \alpha }{{\left( {1 - \theta } \right)}^2}}}{{\left( {1 - 2\beta } \right){\sigma ^2}\left[ {{\theta ^2} + {{\left( {1 - \theta } \right)}^2}} \right]}},\;{\rm{ if }}\;i = A}\\
{\frac{{\beta \eta P\Lambda_Bd_B^{ - \alpha }{\theta ^2}}}{{\left( {1 - 2\beta } \right){\sigma ^2}\left[ {{\theta ^2} + {{\left( {1 - \theta } \right)}^2}} \right]}},\;{\rm{ if }}\;i = B}
\end{array}} \right.$; if $i=A$, $\bar i=B$; if $i=B$, $\bar i=A$.}

\section{Outage Analysis for Dynamic PS Scheme}
In this section, we first propose a dynamic PS scheme to minimize the system outage probability, where the PS ratio for each terminal-relay link is adjusted based on its instantaneous CSI.
Specifically, we obtain the optimal PS ratios in closed forms. Then, we derive the analytical expression for the optimal system outage probability with respect to the optimal PS ratios. Further, the corresponding outage capacity can also be obtained.
\subsection{Dynamic PS Scheme}
Let $P_{\rm{out}}^{s}$ be the overall system outage probability. According to \cite{7831382,6828809}, the system outage probability should jointly consider two E2E outage events and can be defined as the probability that any of the four link data rates  is less than the data rate requirement.
Thus, for a predefined SNR threshold ${\gamma _{\rm{th}}}$, $P_{\rm{out}}^{s}$ is given by
\begin{align}\notag\label{B1}
&P_{\rm{out}}^{s}=1-P_{\rm{success}}^s\\
&=1-\mathbb{P}\left( {{\gamma _{AR}} \geq \gamma _{\rm{th}}}, {{\gamma _{BR}} \geq \gamma _{\rm{th}}}, {{\gamma _{RA}} \geq \gamma _{\rm{th}}},{{\gamma _{RB}} \geq\gamma _{\rm{th}}}\right),
\end{align}
where $P_{\rm{success}}^s$ is the probability that all the four transmissions are successful  and $\mathbb{P} \left(  \cdot  \right)$ denotes the probability.

{\color{black}It is obvious that $P_{\rm{out}}^s$ is always equal to 1 for the cases with $\gamma_{AR}<\gamma_{\rm{th}}$ or $\gamma_{BR}<\gamma_{\rm{th}}$ since the relay can not decode the received signals successfully in such cases.
Thus, in order to achieve the minimum system outage probability, both $\gamma_{AR}\geq\gamma_{\rm{th}}$ and $\gamma_{BR}\geq\gamma_{\rm{th}}$ should be satisfied. For the case with $\gamma_{AR}\geq\gamma_{\rm{th}}$ and $\gamma_{BR}\geq\gamma_{\rm{th}}$,
the values of $\gamma_{RA}$ and $\gamma_{RB}$ decide whether the outage event happens or not for each transmission block, i.e., the lowest system outage probability can be  obtained if we optimize $\rho_A$ and $\rho_B$ to maximize the value of $\min \left( {{\gamma _{RA}},{\gamma _{RB}}} \right)$ for each transmission block. Therefore,
minimizing the system outage probability is equivalent to maximizing the minimum SNR between $\gamma_{RA}$ and $\gamma_{RB}$ as
\begin{align}\label{O1}
\begin{array}{*{20}{l}}
{\mathbf{P1}:\mathop {{\rm{max}}}\limits_{({\rho _A},{\rho _B})} \;\;\min \left( {{\gamma _{RA}},{\gamma _{RB}}} \right)}\\
\begin{array}{l}
{\rm{s}}.{\rm{t}}.\;:\;{\gamma _{AR}} \ge {\gamma _{{\rm{th}}}};\\
\;\;\;\;\;\;\;\;\;\;{\gamma _{BR}} \ge {\gamma _{{\rm{th}}}}.
\end{array}
\end{array}
\end{align}

Based on ${{\gamma _{AR}} \geq \gamma _{\rm{th}}}$ and ${{\gamma _{BR}} \geq \gamma _{\rm{th}}}$, we have ${\rho _i}\leq1 - \frac{{{\gamma _{\rm{th}}}d_i^\alpha {\sigma ^2}}}{{P\Lambda_i{{\left| {{h_i}} \right|}^2}}}$, where $i \in \left\{ {A,B} \right\}$. Thus, the optimization problem $\mathbf{P1}$ can be further rewritten as
\begin{align}\label{O2}
\begin{array}{*{20}{l}}
{\mathbf{P2}:\mathop {{\rm{max}}}\limits_{({\rho _A},{\rho _B})} \;\;\min\left({\gamma _{RA}},{\gamma _{RB}}\right)}\\
{{\rm{s.t.}}\;:\;0 \le {\rho _i} \leq \max\left\{1 - \frac{{{\gamma _{\rm{th}}}d_i^\alpha {\sigma ^2}}}{{P\Lambda_i{{\left| {{h_i}} \right|}^2}}},0\right\}, i \in \left\{ {A,B} \right\}}.\!\!\!
\end{array}
\end{align}}

From \eqref{7}, it is readily seen that ${\gamma _{RA}}\geq {\gamma _{RB}}$ holds for the case with $|{h_A}{|^2}\Lambda_Ad_A^{ - \alpha }{\left( {1 - \theta } \right)^2} \ge |{h_B}{|^2}\Lambda_Bd_B^{ - \alpha }{\theta ^2}$ and that ${\gamma _{RA}}< {\gamma _{RB}}$ is satisfied for the case with $|{h_A}{|^2}\Lambda_Ad_A^{ - \alpha }{\left( {1 - \theta } \right)^2} < |{h_B}{|^2}\Lambda_Bd_B^{ - \alpha }{\theta ^2}$. Since both ${\gamma _{RA}}$ and ${\gamma _{RB}}$ increase with the increase of ${\rho _A}$ and ${\rho _B}$,
the optimal solution to $\mathbf{P2}$ can be obtained when $\rho _A = \max \left\{ {1 - \frac{{\gamma _{\rm{th}}d_A^\alpha {\sigma ^2}}}{{P\Lambda_A{{\left| {{h_A}} \right|}^2}}},0} \right\}$ and $\rho _B = \max \left\{ {1 - \frac{{\gamma _{\rm{th}}d_B^\alpha {\sigma ^2}}}{{P\Lambda_B{{\left| {{h_B}} \right|}^2}}},0} \right\}$.
{\color{black}Thus, the optimal dynamic PS ratio $\rho^{\ast}_{i}, i \in \left\{ {A,B}\right\}$ is given by
\begin{align}\label{9}
\rho _i^* = \max \left\{ {1 - \frac{{\varpi Z_i }}{{{{\left| {{h_i}} \right|}^2}}},0} \right\},
\end{align}
where $\varpi=\frac{{{\gamma _{\rm{th}}}{\sigma ^2}}}{P}$ and $Z_i=\frac{d_i^\alpha}{\Lambda_i}$.}
\subsection{System Outage Probability}
Substituting the optimal PS ratios in \eqref{9} into \eqref{B1}, $P_{\rm{out}}^{s}$ can be expressed as
\begin{align}\label{R2}
P_{\rm{out}}^{s}=1-\mathbb{P}\bigg(|{h_A}{|^2} \ge \Phi_{1},
|{h_B}{|^2} \ge \Phi_{2} \bigg),
\end{align}
where {\color{black}$\Phi_{1}=\max \left(\varpi Z_A {,\frac{{{\gamma _{\rm{th}}}Z_A }}{{{X_B}|{h_B}{|^2}}} + 2\varpi Z_A - |{h_B}{|^2}Z_B^{ - 1 }Z_A } \right)$ and $\Phi_{2}=\max \left( \varpi Z_B{ ,\frac{{{\gamma _{\rm{th}}}Z_B }}{{{X_A}|{h_A}{|^2}}} + 2\varpi Z_B  - |{h_A}{|^2}Z_A^{ - 1 }Z_B} \right)$.}
Based on the values of $\Phi_{1}$ and $\Phi_{2}$, $P_{\rm{out}}^{s}$ can be rewritten as
\begin{align}\label{R3}
P_{\rm{out}}^{s}=1-(P^{s}_{\rm{case1}}+P^{s}_{\rm{case2}}+P^{s}_{\rm{case3}}+P^{s}_{\rm{case4}}),
\end{align}
where $P^{s}_{\rm{case1}} = \mathbb{P}\left(|{h_A}{|^2} \ge \Phi_{1}, |{h_B}{|^2} \ge \Phi_{2}\right)$ with $\Phi_{1}=\frac{{{\gamma _{\rm{th}}}Z_A }}{{{X_B}|{h_B}{|^2}}} + 2\varpi Z_A  - |{h_B}{|^2}Z_B^{ - 1 }Z_A $ and $\Phi_{2}=\varpi Z_B$; $P^{s}_{\rm{case2}}=\mathbb{P}\left(|{h_A}{|^2} \ge \Phi_{1}, |{h_B}{|^2} \ge \Phi_{2}\right)$ with $\Phi_{1}=\varpi Z_A $ and $\Phi_{2}=\frac{{{\gamma _{\rm{th}}}Z_B }}{{{X_A}|{h_A}{|^2}}} + 2\varpi Z_B  - |{h_A}{|^2}Z_A^{ - 1 }Z_B$; $P^{s}_{\rm{case3}}=\mathbb{P}\left(|{h_A}{|^2} \ge \Phi_{1}, |{h_B}{|^2} \ge \Phi_{2}\right)$ with $\Phi_{1}=\varpi Z_A $ and $\Phi_{2}=\varpi Z_B$; and $P^{s}_{\rm{case4}}=\mathbb{P}\left(|{h_A}{|^2} \ge \Phi_{1}, |{h_B}{|^2} \ge \Phi_{2}\right)$ with $\Phi_{1}=\frac{{{\gamma _{\rm{th}}}Z_A }}{{{X_B}|{h_B}{|^2}}} + 2\varpi Z_A - |{h_B}{|^2}Z_B^{ - 1 }Z_A $ and $\Phi_{2}=\frac{{{\gamma _{\rm{th}}}Z_B }}{{{X_A}|{h_A}{|^2}}} + 2\varpi Z_B - |{h_A}{|^2}Z_A^{ - 1 }Z_B$.


Then, the rest of this section is devoted to deriving $P^{s}_{\rm{case1}}$, $P^{s}_{\rm{case2}}$, $P^{s}_{\rm{case3}}$ and $P^{s}_{\rm{case4}}$.

\subsubsection{Derivation  of $P^{s}_{\rm{case1}}$}
Based on the conditions $\Phi_{1}=\frac{{{\gamma _{\rm{th}}}Z_A}}{{{X_B}|{h_B}{|^2}}} + 2\varpi Z_A  - |{h_B}{|^2}Z_B^{ - 1 }Z_A$ and $\Phi_{2}=\varpi Z_B$, the following two equations should be satisfied:
\begin{align}\label{s1}
\left\{ {\begin{array}{*{20}{c}}
{\frac{{{\gamma _{\rm{th}}}}}{{{X_B}}} + \varpi |{h_B}{|^2} - |{h_B}{|^4}Z_B^{ - 1 } \ge 0},\\
{\frac{{{\gamma _{\rm{th}}}}}{{{X_A}}} + \varpi |{h_A}{|^2} - |{h_A}{|^4}Z_A^{ - 1 } < 0}.
\end{array}} \right.
\end{align}
Combining $|{h_A}{|^2}\geq 0$ and $|{h_B}{|^2}\geq 0$, the solution to \eqref{s1} is
$0\leq|{h_B}{|^2}\leq \Delta_{B} $ and $|{h_A}{|^2}>\Delta_{A}$, where $\Delta_{B}=\frac{{\varpi  + \sqrt {{\varpi ^2} + 4{\gamma _{{\rm{th}}}}Z_B^{ - 1}/{X_B}} }}{2}{Z_B}$ and $\Delta_{A}=\frac{{\varpi  + \sqrt {{\varpi ^2} + 4{\gamma _{{\rm{th}}}}Z_A^{ - 1}/{X_A}} }}{2}{Z_A} $.
 Since $\Delta_{B}>\varpi Z_B$ and $\Delta_{A}>\varpi Z_A$, $P^{s}_{\rm{case1}}$ is given by
 \begin{align}\notag\label{s2}
 P^{s}_{\rm{case1}}&=\mathbb{P}\bigg(|{h_A}{|^2} \ge \phi_{A} \left( {|{h_B}{|^2}} \right),\varpi Z_B  \le |{h_B}{|^2} \le {\Delta _B} \bigg)\\
 &\overset{\text{(a)}}{=}\frac{1}{{{\lambda _B}}}\int_{\varpi Z_B }^{{\Delta _B}} {\exp \left( { - \frac{{\phi_{A} \left( x \right)}}{{{\lambda _A}}} - \frac{x}{{{\lambda _B}}}} \right)} dx,
 \end{align}
 where $\phi_{A} \left( {|{h_B}{|^2}} \right) = \max \left( {\frac{{{\gamma _{{\rm{th}}}}Z_A }}{{{X_B}{|{h_B}{|^2}}}} + 2\varpi Z_A  - \frac{{|{h_B}{|^2}}Z_A}{Z_B} ,{\Delta _A}} \right)$ and step (a) holds from $x={{{\left| {{h_B}} \right|}^2}}$ and ${\left| {{h_i}} \right|^2} \sim \exp \left( {\frac{1}{{\lambda _i}}} \right)$ for $i \in \left\{ {A,B} \right\}$.

Since there is no closed-form expression for the integral $\int_{{s_1}}^{{s_2}} {\exp ({z_1}x + \frac{{{z_2}}}{x})} dx$ with any value of $z_{1}$ and $z_{2}\neq 0$, here we employ Gaussian-Chebyshev quadrature \cite{ye11,Neumaier1974Introduction} to  achieve an approximation for $P^{s}_{\rm{case1}}$ as
\begin{small}
\begin{align}\notag\label{s3}
 &P^{s}_{\rm{case1}}\approx \\
 &\frac{{\pi ({\Delta _B} - \varpi Z_B )}}{{2M{\lambda _B}}}\sum\limits_{m = 1}^M  \sqrt {1 - \nu _m^2} \exp \left( { - \frac{{\phi_{A} \left( {\kappa _m^{B}} \right)}}{{{\lambda _A}}} - \frac{{\kappa _m^{B}}}{{{\lambda _B}}}} \right),
\end{align}
\end{small}where $M$ is a parameter that determines the tradeoff between complexity and accuracy
, ${\nu _m} = \cos \frac{{2m - 1}}{{2M}}\pi $, and $\kappa _m^{i} = \frac{{({\Delta _i} - \varpi Z_i )}}{2}{\nu _m} + \frac{{(\varpi Z_i  + {\Delta _i})}}{2}$ for $i \in \left\{ {A,B} \right\}$.

\subsubsection{Derivation of $P^{s}_{\rm{case2}}$}
Based on $\Phi_{1}=\varpi d_A^\alpha $ and $\Phi_{2}=\frac{{{\gamma _{\rm{th}}}Z_B }}{{{X_A}|{h_A}{|^2}}} + 2\varpi Z_B  - |{h_A}{|^2}Z_A^{ - 1 }Z_B$, we have $|{h_B}{|^2} > {\Delta _B}$ and $0 \le |{h_A}{|^2} \le {\Delta _A}$. Then $P^{s}_{\rm{case2}}$ can be calculated as
\begin{small}
\begin{align}\notag\label{s4}
&P^{s}_{\rm{case2}}=\mathbb{P}\bigg(|{h_B}{|^2} \ge {\phi _B}\left( {|{h_A}{|^2}} \right),\varpi Z_A  \le |{h_A}{|^2} \le {\Delta _A}\bigg)\approx\\
&\frac{{\pi ({\Delta _A} - \varpi Z_A )}}{{2M{\lambda _A}}}\sum\limits_{m = 1}^M {\sqrt {1 - \nu _m^2} } \exp \left( { - \frac{{{\phi _B}\left( {\kappa _m^{A}} \right)}}{{{\lambda _B}}} - \frac{{\kappa _m^{A}}}{{{\lambda _A}}}} \right),
\end{align}
\end{small}where ${\phi _B}\left( {|{h_A}{|^2}} \right) = \max \left( {\frac{{{\gamma _{{\rm{th}}}}Z_B }}{{{X_A}|{h_A}{|^2}}} + 2\varpi Z_B  - \frac{{|{h_A}{|^2}Z_B }}{{Z_A }},{\Delta _B}} \right)$.

\subsubsection{Derivation of $P^{s}_{\rm{case3}}$}
Based on $|{h_A}{|^2} \ge \Phi_{1}=\varpi Z_A $ and $|{h_B}{|^2}\ge \Phi_{2}=\varpi Z_B$, the ranges of $|{h_A}{|^2}$ and $|{h_B}{|^2}$ can be given by $|{h_B}{|^2} > {\Delta _B}$ and $|{h_A}{|^2} > {\Delta _A}$, respectively.
Thus $P^{s}_{\rm{case3}}$ is given by
\begin{align}\notag\label{s5}
P^{s}_{\rm{case3}}&=\mathbb{P}\bigg(|{h_A}{|^2} > {\Delta _A}, |{h_B}{|^2} > {\Delta _B} \bigg)\\
&=\exp \left( { - \frac{{{\Delta _A}}}{{{\lambda _A}}} - \frac{{{\Delta _B}}}{{{\lambda _B}}}} \right).
\end{align}

\subsubsection{Derivation of $P^{s}_{\rm{case4}}$}
Based on $\Phi_{1}=\frac{{{\gamma _{\rm{th}}}Z_A }}{{{X_B}|{h_B}{|^2}}} + 2\varpi Z_A  - |{h_B}{|^2}Z_B^{ - 1 }Z_A $ and $\Phi_{2}=\frac{{{\gamma _{\rm{th}}}Z_B}}{{{X_A}|{h_A}{|^2}}} + 2\varpi Z_B  - |{h_A}{|^2}Z_A^{ - 1 }Z_B$, the ranges of $|{h_A}{|^2}$ and $|{h_B}{|^2}$ can be determined by $0\leq|{h_B}{|^2}\leq\Delta_B$ and $0\leq|{h_A}{|^2}\leq\Delta_A$. Thus $P^{s}_{\rm{case4}}$ can be computed as
\begin{align}\label{s51}
P^{s}_{\rm{case4}}=\mathbb{P}\left(\Phi_{1} \le |{h_A}{|^2} \le {\Delta _A},\Phi_{2}\le |{h_B}{|^2} \le {\Delta _B}\right).
\end{align}

Clearly, $\Phi_{1}$ and $\Phi_{2}$ are highly correlated, which is the main difficulty in  deriving $P^{s}_{\rm{case4}}$.
Here, we first determine the integral region that determines the probability $P^{s}_{\rm{case4}}$.
Then we obtain the value of $P^{s}_{\rm{case4}}$ by calculating the integral value over that region.

Let $|{h_A}{|^2} = x$ and $|{h_B}{|^2} = y$.
From the expression of $P^{s}_{\rm{case4}}$ in \eqref{s51}, the integral region of $P^{s}_{\rm{case4}}$ is bounded by 4 lines, which are $y = \frac{{{\gamma _{{\rm{th}}}}Z_B }}{{{X_A}x}} + 2\varpi Z_B  - xZ_A^{ - 1 }Z_B $ (Line 1),
$x = \frac{{{\gamma _{{\rm{th}}}}Z_A }}{{{X_B}y}} + 2\varpi Z_A  - yZ_B^{ - 1 }Z_A $ (Line 2), $x = {\Delta _A}$ (Line 3) and $y = {\Delta _B}$ (Line 4).
Let the intersection points between Line 1 and Line 3 as well as  between Line 1 and  Line 4 be $(x_1,q_1)$ or $(x_\Delta,y_1)$, respectively.

For  $(x_1,q_1)$, the following two equations should be satisfied:
\begin{align}\label{s52}
\left\{ {\begin{array}{*{20}{c}}
{q_1 = \frac{{{\gamma _{{\rm{th}}}}Z_B }}{{{X_A}x_1}} + 2\varpi Z_B  - x_1Z_A^{ - 1 }Z_B},\\
{x_1 = {\Delta _A}}.
\end{array}} \right.
\end{align}
Substituting $x_1={\Delta _A}$ into the first equation of \eqref{s52}, we have $q_1=\frac{{{\gamma _{{\rm{th}}}}Z_B }}{{{X_A}{{\Delta _A}}}} + 2\varpi Z_B  - {{\Delta _A}}Z_A^{ - 1 }Z_B $ and $(x_1,q_1)$ is given by $\left({\Delta _A},\frac{{{\gamma _{{\rm{th}}}}Z_B }}{{{X_A}{{\Delta _A}}}} + 2\varpi Z_B  - {{\Delta _A}}Z_A^{ - 1 }Z_B\right)$. \\

Similarly, for $(x_\Delta,y_1)$, we have
\begin{align}\label{s53}
\left\{ {\begin{array}{*{20}{c}}
{y_1 = \frac{{{\gamma _{{\rm{th}}}}Z_B }}{{{X_A}x_\Delta}} + 2\varpi Z_B  - x_\Delta Z_A^{ - 1 }Z_B},\\
{y_1 = {\Delta _B}}.
\end{array}} \right. .
\end{align}
Further, \eqref{s53} can be rewritten as
\begin{align}\label{s54}
\frac{{{\gamma _{{\rm{th}}}}Z_B }}{{{X_A}}} + \left(2\varpi Z_B-{\Delta _B}\right)x_\Delta  - x_\Delta^2 Z_A^{ - 1 }Z_B=0.
\end{align}
Since $x_\Delta\geq 0$, $x_\Delta$ is given by $\frac{{2\varpi {Z_B} - {\Delta _B} + \sqrt {{{\left( {2\varpi {Z_B} - {\Delta _B}} \right)}^2} + 4{\gamma _{{\rm{th}}}}Z_A^{ - 1}Z_B^2/{X_A}} }}{{2{Z_B}Z_A^{ - 1}}}$. Thus, $(x_\Delta,y_1)$ is given by $\left(\frac{{2\varpi {Z_B} - {\Delta _B} + \sqrt {{{\left( {2\varpi {Z_B} - {\Delta _B}} \right)}^2} + 4{\gamma _{{\rm{th}}}}Z_A^{ - 1}Z_B^2/{X_A}} }}{{2{Z_B}Z_A^{ - 1}}},{\Delta _B}\right)$.

Similarly, the intersection points between Line 2 and Line 3 as well as between Line 1 and  Line 4 are  $(x_1,y_\Delta)$  and  $(q_2,y_1)$ respectively, where $y_\Delta=\frac{{2\varpi {Z_A} - {\Delta _A} + \sqrt {{{\left( {2\varpi {Z_A} - {\Delta _A}} \right)}^2} + 4{\gamma _{{\rm{th}}}}Z_B^{ - 1}Z_A^2/{X_B}} }}{{2{Z_A}Z_B^{ - 1}}}$ and $q_2=\frac{{{\gamma _{{\rm{th}}}}Z_A }}{{{X_B}{{\Delta _B}}}} + 2\varpi Z_A  - {{\Delta _B}}Z_B^{ - 1 }Z_A $.

Let $(x_+,y_+)$ denote the intersection point between Line 1 and Line 2. Then   $(x_+,y_+)$ should satisfy
\begin{align}\label{s6}
\left\{ \begin{array}{l}
x_+= \frac{{{C_A}}}{y_+} + {E_A} - {D_A}y_+,\\
y_+ = \frac{{{C_B}}}{x_+} + {E_B} - {D_B}x_+,
\end{array} \right.
\end{align}
where ${C_i} = \frac{{{\gamma _{{\rm{th}}}}Z_i}}{{{X_{\bar i}}}}$, ${D_i} = Z_{\bar i}^{ - 1 }Z_i $ and ${E_i} = 2\varpi Z_i $. \\

Further, \eqref{s6} can be transformed as
\begin{align}\label{s7}
\left( {{C_A} + {C_B}} \right){x_+^2} - {C_B}{E_A}x_+ - {D_A}C_B^2 = 0.
\end{align}
Since $x_+>0$, the solution to \eqref{s7} is given by $x_+=\frac{{{C_B}{E_A} + \sqrt {C_B^2E_A^2 + 4{D_A}C_B^2\left( {{C_A} + {C_B}} \right)} }}{{2\left( {{C_A} + {C_B}} \right)}}$. Then the corresponding value of $y_+$ is given by $\frac{{{C_B}}}{x_+} + {E_B} - {D_B}x_+$.

Based on the positions of all the intersections, there can be three scenarios for  $P^{s}_{\rm{case4}}$, discussed as follows.

\textbf{Scenario 1:}  When $\max \left( {{q_2},{x_\Delta }} \right) \ge {x_1}$ or $\max \left( {{q_1},{y_\Delta }} \right) \ge {y_1}$, the integral of region for $P^{s}_{\rm{case4}}$ is $0$. Thus, we have $P^{s}_{\rm{case4}}=0$.

\textbf{Scenario 2:} When $\max \left( {{q_2},{x_\Delta }} \right) < {x_1}$, $\max \left( {{q_1},{y_\Delta }} \right) < {y_1}$,  and $x_{+}\leq\max \left( {{q_2},{x_\Delta }} \right)$ (or $x_{+}\geq x_{1}$) hold, the integral of region for $P^{s}_{\rm{case4}}$ is bounded by three lines, which are Line 3, Line 4, and Line 1 (or Line 2).

For the case with $y_\Delta\geq q_1$, the integral of region is bounded by Line 2, Line 3, and Line 4. Then $P^{s}_{\rm{case4}}$ is calculated as
\begin{align}\notag\label{s8}
&P^{s}_{\rm{case4}}=\mathbb{P}\left(\Phi_{1} \le |{h_A}{|^2} \le x_1, y_{\Delta} \le |{h_B}{|^2} \le y_1\right)\\ \notag
&=\frac{1}{{{\lambda _B}}}\exp \left( { - \frac{{{E_A}}}{{{\lambda _A}}}} \right)\int_{{y_\Delta }}^{{y_1}} {\exp \left( {\vartheta _A}\left( y \right) \right)} dy - {f_A}({x_1},{y_\Delta },{y_1})\\ \notag
&\overset{\text{(b)}}{\approx}\frac{{\pi ({y_1} - {y_\Delta })\exp\left( { - \frac{{{E_A}}}{{{\lambda _A}}}} \right)}}{{2M{\lambda _B}}}\sum\limits_{m = 1}^M {\sqrt {1 - \nu _m^2} } \exp \left( {{\vartheta _A}\left( {\kappa _m^{(1)}} \right)} \right) \\
&- {f_A}({x_1},{y_\Delta },{y_1}),
\end{align}
where step (b) holds by using Gaussian-Chebyshev quadrature, ${\vartheta _i}\left( {\kappa _m^{(1)}} \right) =  - \frac{{{C_i}}}{{{\lambda _i}\kappa _m^{(1)}}} + \left( {\frac{{{D_i}}}{{{\lambda _i}}} - \frac{1}{{{\lambda _{\bar i}}}}} \right)\kappa _m^{(1)}$, ${f_i}({x_1},{y_\Delta },{y_1})=\exp \left( { - \frac{{{x_1}}}{{{\lambda _i}}}} \right)\left( {\exp \left( { - \frac{{{y_\Delta }}}{{{\lambda _{\bar i}}}}} \right) - \exp \left( { - \frac{{{y_1}}}{{{\lambda _{\bar i}}}}} \right)} \right)$ and $\kappa _m^{(1)} = \frac{{({y_1} - {y_\Delta })}}{2}{\nu _m} + \frac{{({y_1} + {y_\Delta })}}{2}$.

Likewise, for the case with $y_\Delta< q_1$, the integral of region is bounded
by Line 1, Line 3, and Line 4 and $P^{s}_{\rm{case4}}$ is given by
\begin{align}\notag\label{s9}
&P^{s}_{\rm{case4}}=\mathbb{P}\left(\Phi_{2} \le |{h_B}{|^2} \le y_1, x_{\Delta} \le |{h_A}{|^2} \le x_1\right)\\ \notag
&\approx \frac{{\pi ({x_1} - {x_\Delta })\exp\left( { - \frac{{{E_B}}}{{{\lambda _B}}}} \right)}}{{2M{\lambda _A}}}\sum\limits_{m = 1}^M {\sqrt {1 - \nu _m^2} } \exp \left( {{\vartheta _B}\left( {\kappa _m^{(2)}} \right)} \right) \\
&- {f_B}({y_1},{x_\Delta },{x_1}),
\end{align}
where $\kappa _m^{(2)} = \frac{{({x_1} - {x_\Delta })}}{2}{\nu _m} + \frac{{({x_1} + {x_\Delta })}}{2}$.

\textbf{Scenario 3:} When $\max \left( {{q_2},{x_\Delta }} \right) < {x_1}$, $\max \left( {{q_1},{y_\Delta }} \right) < {y_1}$, and $\max \left( {{q_2},{x_\Delta }} \right)<x_{+}<x_{1}$ are satisfied, the integral region is bounded by 4 lines.
\begin{figure}
  \centering
  \includegraphics[width=0.4\textwidth]{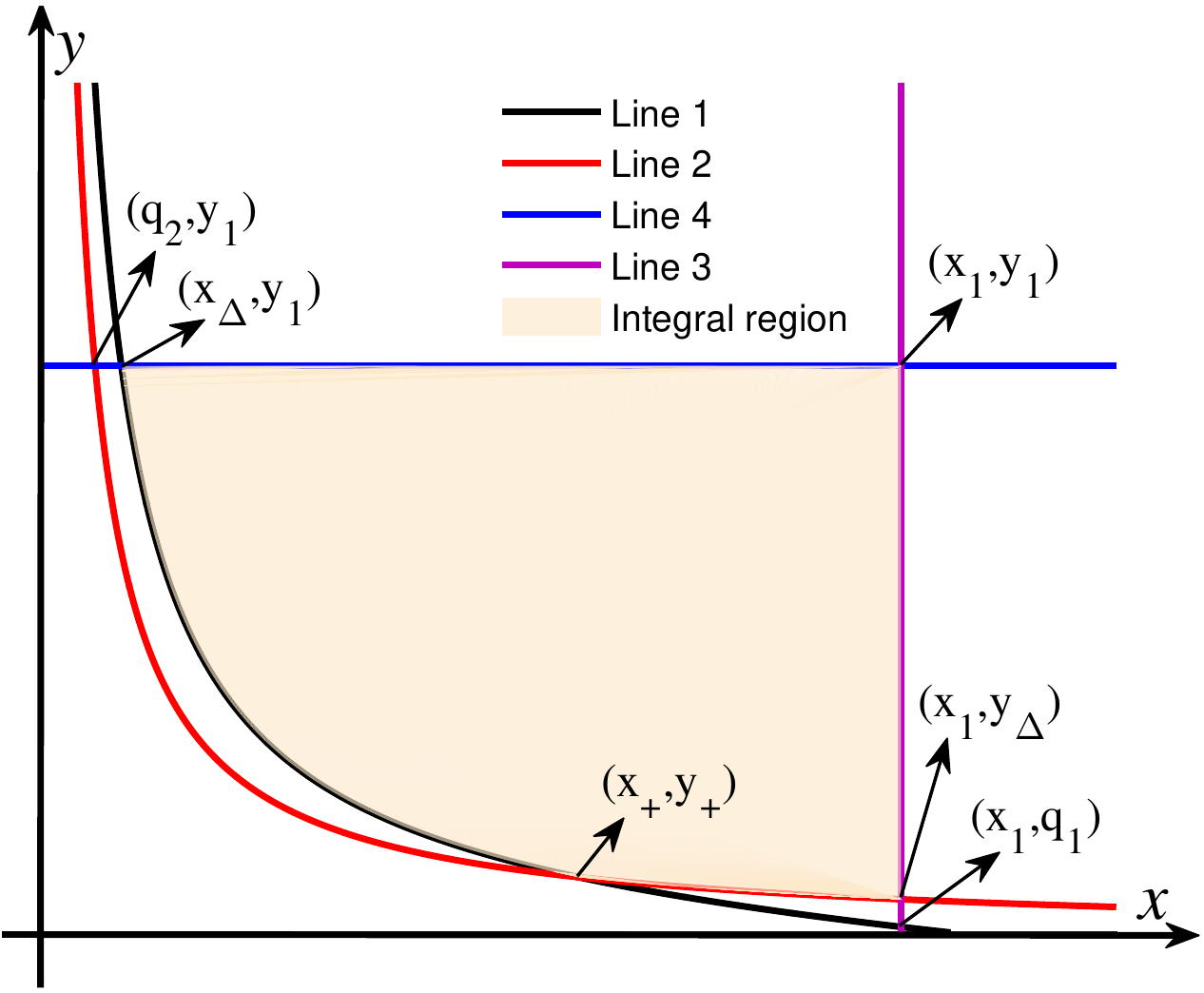}\\
  \caption{Integral region for \textbf{Scenario 3} of  $P^{s}_{\rm{case4}}$  with $y_\Delta\geq q_1$.}\label{fig2}
\end{figure}

For $y_\Delta\geq q_1$, the lower bounds with $x\in[\max \left( {{q_2},{x_\Delta }} \right),x_+]$ and $x\in[x_+,x_1]$ are Line 1 and Line 2, respectively, as shown in Fig. \ref{fig2}.
In this case, $P^{s}_{\rm{case4}}$ can be computed as
\begin{align}\notag\label{s10}
&P^{s}_{\rm{case4}}=\mathbb{P}\left(\Phi_{2} \le |{h_B}{|^2} \le y_1, x_{\Delta} \le |{h_A}{|^2} \le x_+\right)\\ \notag
&+\mathbb{P}\left(\Phi_{1} \le |{h_A}{|^2} \le x_1, y_{\Delta}\le |{h_B}{|^2} \le y_+\right)\\ \notag
&+\mathbb{P}\left(y_+ \le |{h_B}{|^2} \le y_1, x_+ \le |{h_A}{|^2} \le x_1\right)\approx\\ \notag
&
\frac{\pi }{{2M}}\!\!\sum\limits_{m = 1}^M \!\!\!\!{\sqrt {1 - \nu _m^2} }\!\!\!\! \sum\limits_{i = \left\{ {A,B} \right\}}\!\!\!\!\!\! {\frac{{\Delta _{\max }^i - \Delta _{\min }^i}}{{{\lambda _i}}}\exp \left( {{\vartheta _{\bar i}}\left( {\kappa _m^{(i)}} \right) - \frac{{{E_{\bar i}}}}{{{\lambda _{\bar i}}}}} \right)}+\\
&\Xi_A(x_+,x_1,y_+,y_1)\!-\! {f_B}({y_1},{x_\Delta },{x_ + })\! - \! {f_A}({x_1},{y_\Delta },{y_ + }),\!\!\!\!
\end{align}
where $\Delta _{\max }^i$ is determined by $\left\{ {\begin{array}{*{20}{c}}
{{x_ + },{\rm{ }}i = A}\\
{{y_ + },{\rm{ }}i = B}
\end{array}} \right.$; $\Delta _{\min }^i$ is given by $\left\{ {\begin{array}{*{20}{c}}
{{x_\Delta },{\rm{ }}i = A}\\
{{y_\Delta },{\rm{ }}i = B}
\end{array}} \right.$; $\kappa _m^{(i)}$ is given by $\frac{{(\Delta _{\max }^i - \Delta _{\min }^i)}}{2}{\nu _m} + \frac{{(\Delta _{\max }^i + \Delta _{\min }^i)}}{2}$; and
$\Xi_i(x_+,x_1,y_+,y_1)=\left( {\exp \left( { - \frac{{{x_ + }}}{{{\lambda _i}}}} \right) - \exp \left( { - \frac{{{x_1}}}{{{\lambda _i}}}} \right)} \right)\left( {\exp \left( { - \frac{{{y_ + }}}{{{\lambda _{\bar i}}}}} \right) - \exp \left( { - \frac{{{y_1}}}{{{\lambda _{\bar i}}}}} \right)} \right)$.

For $y_\Delta< q_1$, the lower bounds with $x\in[\max \left( {{q_2},{x_\Delta }} \right),x_+]$ and $x\in[x_+,x_1]$ are Line 2 and Line 1, respectively.
In this case, $P^{s}_{\rm{case4}}$ is given by
\begin{align}\notag\label{s10}
&P^{s}_{\rm{case4}}\approx - {f_B}({y_+},{x_+ },{x_ 1 }) - {f_A}({x_1},{y_+ },{y_ 1 })+\\
&\frac{\pi }{{2M}}\!\!\sum\limits_{m = 1}^M \!\!\!{\sqrt {1 - \nu _m^2} } \!\!\!\! \sum\limits_{i = \left\{ {A,B} \right\}}\!\!\!\!\!\! {\frac{{\omega _{\max }^i - \omega _{\min }^i}}{{{\lambda _i}}}\exp \left( {{\vartheta _{\bar i}}\left( {\kappa _m^{[i]}} \right) - \frac{{{E_{\bar i}}}}{{{\lambda _{\bar i}}}}} \right)},
\end{align}
where $\omega _{\max }^i$ is given by $\left\{ {\begin{array}{*{20}{c}}
{{x_1},{\rm{ }}i = A}\\
{{y_1},{\rm{ }}i = B}
\end{array}} \right.$;  $\omega _{\min }^i = \left\{ {\begin{array}{*{20}{c}}
{{x_ + },{\rm{ }}i = A}\\
{{y_ + },{\rm{ }}i = B}
\end{array}} \right.$; and $\kappa _m^{[i]}$ is given by $\frac{{(\omega _{\max }^i - \omega _{\min }^i)}}{2}{\nu _m} + \frac{{(\omega_{\max }^i + \omega _{\min }^i)}}{2}$.

\begin{figure*}
\begin{align}\label{G3}
&P_{\rm{out}}^{s}\approx1-\exp \left( { - \frac{{{\Delta _A}}}{{{\lambda _A}}} - \frac{{{\Delta _B}}}{{{\lambda _B}}}} \right)-\frac{\pi }{{2M}}\sum\limits_{m = 1}^M {\sqrt {1 - \nu _m^2} } \sum\limits_{i = \left\{ {A,B} \right\}} {\frac{{{\Delta _i} - \varpi Z_i }}{{{\lambda _i}}}\exp \left( { - \frac{{{\phi _{\bar i}}\left( {\kappa _m^i} \right)}}{{{\lambda _{\bar i}}}} - \frac{{\kappa _m^i}}{{{\lambda _i}}}} \right)}-\Theta,\\ \notag
\end{align}
where
\begin{align}\notag
&\Theta  = \left\{ {\begin{array}{*{20}{c}}
{\!\!\!\!\!\!\!\!\!\!\!\!\!\!\!\!0,{\rm{   \; Scenario\; 1;}}}\\
\!\!\!\!\!\!\!\!\!\!\!\!\!\!\!\!\!\!\!\!\!\!\!\!\!\!\!\!\!\!\!\!\!\!\!\!\!\!\!\!\!\!\!{\frac{{\pi ({y_1} - {y_\Delta })\exp \left( { - \frac{{{E_A}}}{{{\lambda _A}}}} \right)}}{{2M{\lambda _B}}}\sum\limits_{m = 1}^M {\sqrt {1 - \nu _m^2} } \exp \left( {{\vartheta _A}\left( {\kappa _m^{(1)}} \right)} \right) - {f_A}({x_1},{y_\Delta },{y_1}),\;{\rm{   Scenario \;2\; with \;}}{y_\Delta } \ge {q_1}};\\
\!\!\!\!\!\!\!\!\!\!\!\!\!\!\!\!\!\!\!\!\!\!\!\!\!\!\!\!\!\!\!\!\!\!\!\!\!\!\!\!\!\!\!\!\!{\frac{{\pi ({x_1} - {x_\Delta })\exp \left( { - \frac{{{E_B}}}{{{\lambda _B}}}} \right)}}{{2M{\lambda _A}}}\sum\limits_{m = 1}^M {\sqrt {1 - \nu _m^2} } \exp \left( {{\vartheta _B}\left( {\kappa _m^{(2)}} \right)} \right) - {f_B}({y_1},{x_\Delta },{x_1}),{\rm{  \; Scenario\; 2 \;with\; }}{y_\Delta } < {q_1}};\\
\begin{array}{l}
\!\!\!\!\!\!{\Xi _A}({x_ + },{x_1},{y_ + },{y_1}) + \frac{\pi }{{2M}}\sum\limits_{m = 1}^M {\sqrt {1 - \nu _m^2} } \sum\limits_{i = \left\{ {A,B} \right\}} {\frac{{\Delta _{\max }^i - \Delta _{\min }^i}}{{{\lambda _i}}}\exp \left( {{\vartheta _{\bar i}}\left( {\kappa _m^{(i)}} \right) - \frac{{{E_{\bar i}}}}{{{\lambda _{\bar i}}}}} \right)}  - {f_B}({y_1},{x_\Delta },{x_ + }) - {f_A}({x_1},{y_\Delta },{y_ + }),\\
\;\;\;\;\;\;\;\;\;\;\;\;\;\;\;\;\;\;\;\;\;\;\;\;\;\;\;\;\;\;\;\;\;\;\;\;\;\;\;\;\;\;\;\;\;\;\;\;\;\;\;\;\;\;\;\;\;\;\;\;\;\;\;\;\;\;\;\;\;\;\;\;\;\;\;\;\;\;\;\;\;\;\;\;\;\;\;\;\;\;\;\;\;\;\;\;\;\;\;\;\;\;\;\;\;\;\;\;\;\;\;\;\;\;\;\;\;\;\;\;\;\;\;\;\;\;\;\;\;\;\;\;\;\;\;\;\;\;\;\;{\rm{  Scenario\; 3 \;with \;}}{y_\Delta } \ge {q_1};\\
\!\!\!\!\!\!\frac{\pi }{{2M}}\sum\limits_{m = 1}^M {\sqrt {1 - \nu _m^2} } \sum\limits_{i = \left\{ {A,B} \right\}} {\frac{{\omega _{\max }^i - \omega _{\min }^i}}{{{\lambda _i}}}\exp \left( {{\vartheta _{\bar i}}\left( {\kappa _m^{[i]}} \right) - \frac{{{E_{\bar i}}}}{{{\lambda _{\bar i}}}}} \right)}  - {f_B}\left( {{y_ + },{x_ + },{x_1}} \right) - {f_A}\left( {{x_1},{y_ + },{y_1}} \right),{\rm{\;  Scenario \; 3\; with \;}}{y_\Delta } < {q_1}.{\rm{ }}
\end{array}
\end{array}} \right.
.\end{align}
\hrulefill
\end{figure*}

Combining \eqref{R3} with $P^{s}_{\rm{case1}}$, $P^{s}_{\rm{case2}}$, $P^{s}_{\rm{case3}}$ and $P^{s}_{\rm{case4}}$, the value of $P_{\rm{out}}^{s}$ can be determined in \eqref{G3} at the top of next page.

{\color{black}\textit{Remark:} The derived expression for system outage probability in \eqref{G3} serves the following purposes. First, \eqref{G3} can characterize the system outage probability of SWIPT enabled three-step two-way DF relay networks for the dynamic PS scheme with a small $M$ and certain accuracy instead of carrying out computer simulations. Second,  we can obtain some insightful understandings on selecting proper system parameters based on the curves obtained by \eqref{G3}. Note that this approach has also adopted in many works, e.g., \cite{AA13,7565756,2017CL}. Third, based on the derived expression in \eqref{G3}, we can compute the system outage capacity and analyze the diversity gain for our investigated network, as shown in Section III.C and Section III.D.}

\subsection{System Outage Capacity}
Based on the analytical result of the system outage probability in \eqref{G3}, we can obtain the system outage capacity for the dynamic PS scheme, denoted by $\tau^{\rm{DPS}}$. Since both $A$ and $B$ transmit signals at the transmission rate $U=\log_{2}(1+\gamma_{\rm{th}})$, and the effective transmission time is given by the minimum of $\beta T$ and $(1-2\beta)T$, the system outage capacity $\tau^{\rm{DPS}}$ is given by
\begin{align}\label{G11}
\tau^{\rm{DPS}}=(1-P_{\rm{out}}^{s})U\times\min\left(\beta T, (1-2\beta)T\right).
\end{align}

{\color{black}
\subsection{Diversity Gain}
According to \cite{ye11,7032100}, the diversity gain of the investigated system under the dynamic PS scheme can be computed as
\begin{align}\label{b1}\notag
&d =  - \mathop {\lim }\limits_{{\rho _0} \to \infty } \frac{{\log \left( P_{\rm{out}}^{s} \right)}}{{\log \left( {{\rho _0}} \right)}}\\
&=\!- \!\mathop {\lim }\limits_{{\rho _0} \to \infty } \frac{{\log \left( {1 - P^{s}_{\rm{case1}} - P^{s}_{\rm{case2}} - P^{s}_{\rm{case3}} - P^{s}_{\rm{case4}}} \right)}}{{\log \left( {{\rho _0}} \right)}},\!\!\!
\end{align}
where $\rho_0=\frac{P}{\sigma^2}$ denotes the input SNR.

Based on the expression of $P^{s}_{\rm{case1}}$, we have $\mathop {\lim }\limits_{{\rho _0} \to \infty } P^{s}_{\rm{case1}}=0$ due to the fact that $\mathop {\lim }\limits_{{\rho _0} \to \infty } \Delta_B=0=\mathop {\lim }\limits_{{\rho _0} \to \infty } \varpi Z_B$. Likewise, $\mathop {\lim }\limits_{{\rho _0} \to \infty } P^{s}_{\rm{case2}}$ is given by 0 since  $\mathop {\lim }\limits_{{\rho _0} \to \infty } \Delta_A=0=\mathop {\lim }\limits_{{\rho _0} \to \infty } \varpi Z_A$. Since $\mathop {\lim }\limits_{{\rho _0} \to \infty } x_1=\mathop {\lim }\limits_{{\rho _0} \to \infty } y_1=\mathop {\lim }\limits_{{\rho _0} \to \infty } x_\Delta=\mathop {\lim }\limits_{{\rho _0} \to \infty } y_\Delta=\mathop {\lim }\limits_{{\rho _0} \to \infty } q_2=\mathop {\lim }\limits_{{\rho _0} \to \infty } q_1=0$, we have $\mathop {\lim }\limits_{{\rho _0} \to \infty } P^{s}_{\rm{case4}}=0$. Then the diversity gain can be rewritten as
\begin{align}\label{b2}\notag
d &=  - \mathop {\lim }\limits_{{\rho _0} \to \infty } \frac{{\log \left( {1 - P^{s}_{\rm{case3}}} \right)}}{{\log \left( {{\rho _0}} \right)}}\\ \notag
&=-\mathop {\lim }\limits_{{\rho _0} \to \infty } \frac{{\log \left( {1 - \exp \left( { - \frac{1}{{{\rho _0}}}} \right)} \right)}}{{\log \left( {{\rho _0}} \right)}}\\
&\mathop {{\rm{  }} = }\limits^{x = \frac{1}{{{\rho _0}}}} \mathop {\lim }\limits_{x \to 0} \frac{x}{{1 - \exp \left( { - x} \right)}} = 1.
\end{align}}

\section{Outage Analysis for Improved Dynamic Scheme}
In this section, by considering the combining strategy at the relay, we further develop an improved dynamic scheme to improve the system outage performance.
For the improved dynamic scheme, we jointly optimize the PS ratios and power allocation ratio $\theta$ used for combining the decoded signals at the relay to achieve the minimum system outage probability.
In particular, we first find the optimal values for PS ratios and power allocation ratio $\theta$, respectively. On this basis, we derive an analytical expression for the optimal system outage probability and obtain the optimal system outage capacity.

\subsection{Improved Dynamic Scheme}
Specifically, the optimization problem to minimize the system outage problem is formulated as
\begin{align}\label{s11}
\begin{array}{*{20}{l}}
{\mathbf{P3}:\mathop {{\rm{maximize}}}\limits_{({\rho _A},{\rho _B},\theta )} \;\;\min \left( {{\gamma _{RA}},{\gamma _{RB}}} \right)}\\
\begin{array}{l}
{\rm{s}}.{\rm{t}}.\;:\;0 \le {\rho _i} \le \max \left\{ {1 - \frac{{\varpi Z_i }}{{{{\left| {{h_i}} \right|}^2}}},0} \right\},i \in \left\{ {A,B} \right\},\\
\;\;\;\;\;\;\;\;\;\;{\rm{       }}0 < \theta  < 1.
\end{array}
\end{array}
\end{align}

Clearly, for any given $\theta$, the optimal PS ratios is given by ${\rho _i^*}, i \in \left\{ {A,B} \right\}$ in \eqref{9} according to the expressions of ${\gamma _{RA}}$ and ${\gamma _{RB}}$.

Substituting the optimal PS ratios into ${\gamma _{RA}}$ and ${\gamma _{RB}}$, $\mathbf{P3}$ can be reformulated as
\begin{align}\label{s111}
\begin{array}{*{20}{l}}
{\mathbf{P4}:\mathop {{\rm{maximize}}}\limits_\theta  \;\;\min \left(\frac{{{\Omega _1}{{\left( {1 - \theta } \right)}^2}}}{{{\theta ^2} + {{\left( {1 - \theta } \right)}^2}}}, \frac{{{\Omega _2}{\theta ^2}}}{{{\theta ^2} + {{\left( {1 - \theta } \right)}^2}}} \right)}\\
{{\rm{s}}.{\rm{t}}.\;:\;0 < \theta  < 1,}
\end{array}
\end{align}
where ${\Omega _1} = \frac{{\eta \beta PZ_A^{ - 1 }}}{{\left( {1 - 2\beta } \right){\sigma ^2}}}\left( {\rho _A^*|{h_A}{|^4}Z_A^{ - 1 } + \rho _B^*|{h_B}{|^2}|{h_A}{|^2}Z_B^{ - 1 }} \right)$ and ${\Omega _2} = \frac{{\eta \beta PZ_B^{ - 1 }}}{{\left( {1 - 2\beta } \right){\sigma ^2}}}\left( {\rho _B^*|{h_B}{|^4}Z_B^{ - 1 } + \rho _A^*|{h_B}{|^2}|{h_A}{|^2}Z_A^{ - 1 }} \right)$.

Similarly, based on the relation of $\frac{{{\Omega _1}{{\left( {1 - \theta } \right)}^2}}}{{{\theta ^2} + {{\left( {1 - \theta } \right)}^2}}}$ and $\frac{{{\Omega _2}{\theta ^2}}}{{{\theta ^2} + {{\left( {1 - \theta } \right)}^2}}}$, the optimization problem $\mathbf{P4}$ can be divided into two scenarios: \textbf{Scenario I:} $ \frac{{{\Omega _1}{{\left( {1 - \theta } \right)}^2}}}{{{\theta ^2} + {{\left( {1 - \theta } \right)}^2}}}\geq \frac{{{\Omega _2}{\theta ^2}}}{{{\theta ^2} + {{\left( {1 - \theta } \right)}^2}}}$ and \textbf{Scenario II:} $ \frac{{{\Omega _1}{{\left( {1 - \theta } \right)}^2}}}{{{\theta ^2} + {{\left( {1 - \theta } \right)}^2}}}\leq \frac{{{\Omega _2}{\theta ^2}}}{{{\theta ^2} + {{\left( {1 - \theta } \right)}^2}}}$.

\textbf{Scenario I:} Based on $ \frac{{{\Omega _1}{{\left( {1 - \theta } \right)}^2}}}{{{\theta ^2} + {{\left( {1 - \theta } \right)}^2}}}\geq \frac{{{\Omega _2}{\theta ^2}}}{{{\theta ^2} + {{\left( {1 - \theta } \right)}^2}}}$, we have
$|{h_A}{|^2}Z_A^{ - 1 }{\left( {1 - \theta } \right)^2} \ge |{h_B}{|^2}Z_B^{ - 1 }{\theta ^2}$. In this scenario, $\mathbf{P4}$ can be rewritten as
\begin{align}
\begin{array}{*{20}{l}}
{\mathbf{P4a}:\mathop {{\rm{maximize}}}\limits_\theta  \;\;\frac{{{\Omega _2}{\theta ^2}}}{{{\theta ^2} + {{\left( {1 - \theta } \right)}^2}}}}\\
{{\rm{s}}.{\rm{t}}.\;:\;0 < \theta  \le \frac{{|{h_A}|\sqrt{Z_B}}}{{|{h_A}|\sqrt{Z_B} + |{h_B}|\sqrt{Z_A}}}.}
\end{array}
\end{align}

Since $\frac{{{\Omega _2}{\theta ^2}}}{{{\theta ^2} + {{\left( {1 - \theta } \right)}^2}}}$ is a monotonic increasing function of $\theta$, the optimal solution to $\mathbf{P4a}$ is given by $\frac{{|{h_A}|\sqrt{Z_B}}}{{|{h_A}|\sqrt{Z_B} + |{h_B}|\sqrt{Z_A}}}$.

\textbf{Scenario II:} Similarly, according to $ \frac{{{\Omega _1}{{\left( {1 - \theta } \right)}^2}}}{{{\theta ^2} + {{\left( {1 - \theta } \right)}^2}}}\leq \frac{{{\Omega _2}{\theta ^2}}}{{{\theta ^2} + {{\left( {1 - \theta } \right)}^2}}}$, $\mathbf{P4}$ can be given by
\begin{align}
\begin{array}{*{20}{l}}
{\mathbf{P4b}:\mathop {{\rm{maximize}}}\limits_\theta  \;\;\frac{{{\Omega _1}{{\left( {1 - \theta } \right)}^2}}}{{{\theta ^2} + {{\left( {1 - \theta } \right)}^2}}}}\\
{{\rm{s}}.{\rm{t}}.\;:\;\frac{{|{h_A}|\sqrt{Z_B}}}{{|{h_A}|\sqrt{Z_B} + |{h_B}|\sqrt{Z_A}}}\leq \theta <1 .}
\end{array}
\end{align}

Since $\frac{{{\Omega _1}{{\left( {1 - \theta } \right)}^2}}}{{{\theta ^2} + {{\left( {1 - \theta } \right)}^2}}}=\frac{{{\Omega _1}}}{{{{\left( {\frac{1}{{{1 \mathord{\left/
 {\vphantom {1 {\theta  - 1}}} \right.
 \kern-\nulldelimiterspace} {\theta  - 1}}}}} \right)}^2} + 1}}$ decreases with the increasing of $\theta$, the optimal solution is also given by $\frac{{|{h_A}|\sqrt{Z_B}}}{{|{h_A}|\sqrt{Z_B} + |{h_B}|\sqrt{Z_A}}}$.

In summary, the optimal solutions to $\mathbf{P4}$ are given by
\begin{align}
&\rho _i^* = \max \left\{ {1 - \frac{{\varpi Z_i }}{{{{\left| {{h_i}} \right|}^2}}},0} \right\}, i \in \left\{ {A,B}\right\}, \\
&\theta^*= \frac{{|{h_A}|\sqrt{Z_B}}}{{|{h_A}|\sqrt{Z_B} + |{h_B}|\sqrt{Z_A}}}.
\end{align}

\subsection{System Outage Probability}
Substituting ${\rho _i^*}$ and $\theta^*$ into \eqref{B1}, the optimal system outage probability for the improved dynamic scheme is given by
\begin{align}\notag\label{s13}
&P_{\rm{out}}^{ss}=1-\mathbb{P}\bigg(|{h_A}{|^2} \ge \varpi Z_A ,|{h_B}{|^2} \ge \varpi Z_B,\\ \notag
&Y\frac{{|{h_B}{|^2}|{h_A}{|^4}Z_A^{ - 1 } + |{h_B}{|^4}|{h_A}{|^2}Z_B^{ - 1 } - 2\varpi |{h_B}{|^2}|{h_A}{|^2}}}{{|{h_A}{|^2}Z_A^{ - 1 } + |{h_B}{|^2}Z_B^{ - 1 }}} \ge {\gamma _{{\rm{th}}}}\bigg)\\
&\overset{\text{(c)}}{=}1-\mathbb{P}\left({t_2}\left( {1 - \frac{{{\gamma _{{\rm{th}}}}}}{Y}{t_3}} \right) \ge 2\varpi ,{t_2} \le \frac{1}{{\varpi Z_A Z_B {t_3}}} + \varpi,{t_2} \ge 2\varpi \right)\!\!\!\!\!\!
\end{align}
where $Y = \frac{{\eta \beta PZ_A^{ - 1 }Z_B^{ - 1 }}}{{\left( {1 - 2\beta } \right){\sigma ^2}}}$ and step (c) holds by letting ${t_2} = |{h_A}{|^2}Z_A^{ - 1 } + |{h_B}{|^2}Z_B^{ - 1 }$ and ${t_3} = \frac{1}{{|{h_A}{|^2}|{h_B}{|^2}}}$.

Note that when ${1 - \frac{{{\gamma _{{\rm{th}}}}}}{Y}{t_3}}\leq 0$ is satisfied, $0<2\varpi\leq t_2<0$ can be obtained and $P_{\rm{out}}^{ss}$ in this case is $0$.
Thus, the value of $P_{\rm{out}}^{ss}$ is equal to the value of $P_{\rm{out}}^{ss}$ with ${1 - \frac{{{\gamma _{{\rm{th}}}}}}{Y}{t_3}}> 0$. Besides, by letting $\frac{{2\varpi }}{{1 - \frac{{{\gamma _{{\rm{th}}}}}}{Y}{t_3}}}\le \frac{1}{{\varpi Z_A Z_B {t_3}}} + \varpi$, we have
\begin{align}\label{sss}
a_ot_3^2 + b_o{t_3} - 1 \le 0
\end{align}
where $a_o=\frac{{{\varpi ^2}Z_A Z_B {\gamma _{{\rm{th}}}}}}{Y}$ and $b_o={{\varpi ^2}Z_A Z_B  + \frac{{{\gamma _{{\rm{th}}}}}}{Y}}$. Then the range of $t_3$ is given by $0\leq t_3\leq\frac{{\sqrt {b_o^2 + 4{a_o}}  - {b_o}}}{{2{a_o}}}$.

Thus, $P_{\rm{out}}^{ss}$ can be rewritten as
\begin{align}\notag\label{s14}
&P_{\rm{out}}^{ss}=1-\\
&\mathbb{P}\left(\frac{{2\varpi }}{{1 - \frac{{{\gamma _{{\rm{th}}}}}}{Y}{t_3}}}\leq {t_2} \le \frac{1}{{\varpi Z_A Z_B {t_3}}} + \varpi ,0 \le {t_3} \le {t_{\max }} \right)
\end{align}
where ${t_{\max }} = \min \left( \frac{{\sqrt {b_o^2 + 4{a_o}}  - {b_o}}}{{2{a_o}}},\frac{Y}{{{\gamma _{{\rm{th}}}}}} \right)$.
Further, the following \textbf{Lemma. 1} is provided to derive $P_{\rm{out}}^{ss}$ in \eqref{s14}.

\textbf{Lemma. 1} The cumulative distribution functions (CDFs) of $t_2$ and $t_3$ are given by
\begin{align}\notag\label{s15}
&F_{t_2}(t)=\\
&\left\{ \begin{array}{l}
\!\!\!\!1 - {e^{ - {a_B}t}} - \frac{{{a_B}}}{{{a_A} - {a_B}}}\left( {{e^{{- {a_B}} t}} - {e^{ { - {a_A}} t}}} \right),{\rm{ if}}\;{a_A} \ne {a_B},\\
\!\!\!\!1 - {e^{ - {a_B}t}} - {a_B}t{e^{ { - {a_A}} t}},{\rm{ if }}\;{a_A} = {a_B};
\end{array} \right.    \\
&F_{t_3}(t)=\frac{1}{{{\lambda _B}}}\sqrt {\frac{{4{\lambda _B}}}{{{\lambda _A}t}}} {K_1}\left( {\sqrt {\frac{4}{{{\lambda _A}{\lambda _B}t}}} } \right);
\end{align}
where ${a_i} = \frac{{Z_i }}{{{\lambda _i}}}$ and ${K_1}\left(  \cdot  \right)$ is the modified Bessel function of the second kind.

\emph{Proof:} See the Appendix. \hfill {$\blacksquare $}

Let ${{f_{{t_2}}}\left( {{t}} \right)}$ and ${f_{{t_3}}}\left( {{t}} \right)$ denote the probability density functions (PDF) of $F_{t_2}(t)$ and $F_{t_3}(t)$, respectively, and we have ${{f_{{t_2}}}\left( {{t}} \right)}=\frac{{\partial {F_{{t_2}}}\left( t \right)}}{{\partial t}}$ and ${f_{{t_3}}}\left( {{t}} \right)=\frac{{\partial {F_{{t_3}}}\left( t \right)}}{{\partial t}}$.

Then $P_{\rm{out}}^{ss}$ can be calculated as
\begin{align}\notag\label{s161}
P_{\rm{out}}^{ss}&=1 - \int_0^{{t_{\max }}} {\int_{\frac{{2\varpi }}{{1 - \frac{{{\gamma _{{\rm{th}}}}}}{Y}{t_3}}}}^{ \frac{1}{{\varpi Z_A Z_B {t_3}}} + \varpi } {{f_{{t_2}}}\left( {{t_2}} \right)} } {f_{{t_3}}}\left( {{t_3}} \right)d{t_2}d{t_3}\\
&=1 - \int_0^{{t_{\max }}} {\chi ({t_3})} {f_{{t_3}}}\left( {{t_3}} \right)d{t_3},
\end{align}
where $\chi(t_3)={F_{t_2}\left(\frac{1}{{\varpi Z_A Z_B {t_3}}} + \varpi\right) - {F_{{t_2}}}\left( {\frac{{2\varpi }}{{1 - \frac{{{\gamma _{{\rm{th}}}}}}{Y}t_3}}} \right)}$.

Using the subsection integral method, $P_{\rm{out}}^{ss}$ can be further calculated as
\begin{align}\notag\label{s16}
P_{\rm{out}}^{ss}
&=1-\chi(t){F_{{t_3}}}\left( t \right)|^{{t_{\max }}}_{0}+\int_0^{{t_{\max }}} {\chi '(t)} {F_{{t_3}}}\left( t \right)dt\\
&=1-\chi(t_{\max }){F_{{t_3}}}\left( t_{\max } \right)+\int_0^{{t_{\max }}} {\chi '(t)} {F_{{t_3}}}\left( t \right)dt,
\end{align}
where 
\begin{align}\notag
&\chi '(t) = \frac{{\partial \chi (t)}}{{\partial t}}=\\ \notag
&\left\{ {\begin{array}{*{20}{l}}
{\frac{{{a_A}{a_B}}}{{{a_A} - {a_B}}}{{\sum\limits_{j = 1}^2\!\!{\left( { - 1} \right)} }^{j + 1}}\!\!\!\!\!\!\!{{s'}_j}\left( t \right)\left( {e^{ - {a_B}{s_j}\left( t \right)}- e^{ - {a_A}{s_j}\left( t \right)}} \right),{\rm{if}}\;{a_A} \ne {a_B}}\\
{a_A^2{{\sum\limits_{j = 1}^2 {\left( { - 1} \right)} }^{j + 1}}{s_j}\left( t \right){{s'}_j}\left( t \right)e^{ - {a_A}{s_j}\left( t \right)},{\rm{if}}\;{a_A} = {a_B}}
\end{array}} \right.
\end{align}
with ${s_1}\left( t \right) = \frac{1}{{\varpi Z_A Z_B t}} + \varpi $, $s_{2}\left( t \right) = \frac{{2\varpi }}{{1 - \frac{{{\gamma _{{\rm{th}}}}}}{Y}t}}$, ${{s'}_1}\left( t \right) =  - \frac{1}{{\varpi Z_A Z_B {t^2}}}$ and $s'_{2}\left( t \right) = \frac{{2\varpi \frac{{{\gamma _{{\rm{th}}}}}}{Y}}}{{{{\left( {1 - \frac{{{\gamma _{{\rm{th}}}}}}{Y}t} \right)}^2}}}$. 

By using Gaussian-Chebyshev quadrature, $P_{\rm{out}}^{ss}$ can be approximated as
\begin{align}\notag\label{s17}
&P_{\rm{out}}^{ss}\approx 1-\chi(t_{\max }){F_{{t_3}}}\left( t_{\max } \right)\\
&+\frac{{\pi {t_{\max }}}}{{2M}}\sum\limits_{m = 1}^M {\sqrt {1 - \nu _m^2} } \chi '(\kappa _m^{(3)}){F_{{t_3}}}\left( {\kappa _m^{(3)}} \right),
\end{align}
where $\kappa _m^{(3)} = \frac{{{t_{\max }}}}{2}{\nu _m} + \frac{{{t_{\max }}}}{2}$.

\subsection{System Outage Capacity}
In this subsection, we achieve the approximation of the system outage capacity for the improved dynamic scheme based on the expression of the system outage probability in \eqref{s17}. Let $\tau^{\rm{IDS}}$ denote the system outage capacity. Then $\tau^{\rm{IDS}}$ can be computed as
\begin{align}\notag\label{G111}
\tau^{\rm{IDS}}&=(1-P_{\rm{out}}^{ss})U\times\min\left(\beta T, (1-2\beta)T\right)\\ \notag
&\approx \bigg(- \frac{{\pi {t_{\max }}}}{{2M}}\sum\limits_{m = 1}^M {\sqrt {1 - \nu _m^2} } \chi '(\kappa _m^{(3)}){F_{{t_3}}}\left( {\kappa _m^{(3)}} \right)\\
&+\chi ({t_{\max }}){F_{{t_3}}}\left( {{t_{\max }}} \right)\bigg)UT\times\min\left(\beta, 1-2\beta\right).\!\!
\end{align}

{\color{black}
\subsection{Diversity Gain}
The diversity gain of the investigated system under the improved dynamic scheme can be calculated as
\begin{align}\label{b3}\notag
d &=  - \mathop {\lim }\limits_{{\rho _0} \to \infty } \frac{{\log \left( P_{\rm{out}}^{ss} \right)}}{{\log \left( {{\rho _0}} \right)}}=- \mathop {\lim }\limits_{{\rho _0} \to \infty } \frac{{\log \left( 1 - {F_{{t_3}}}\left( t_{\max } \right) \right)}}{{\log \left( {{\rho _0}} \right)}}\\ \notag
\!\!\!\!\!\!&\mathop {{\rm{}} = }\limits^{x = \frac{1}{{{\rho _0}}}}- \mathop {\lim }\limits_{x \to 0} \frac{{\log \left( {1 - \sqrt x {K_1}\left( {\sqrt x } \right)} \right)}}{{\log \left( {{1 \mathord{\left/
 {\vphantom {1 x}} \right.
 \kern-\nulldelimiterspace} x}} \right)}}\\ \notag
 &\overset{\text{(a)}}{=}- \mathop {\lim }\limits_{x \to 0}\frac{{\log \left(- {\frac{x}{2}\left( {\ln\frac{{\sqrt x }}{2} + {c_0}} \right)} \right)}}{{\log \left( {{1 \mathord{\left/
 {\vphantom {1 x}} \right.
 \kern-\nulldelimiterspace} x}} \right)}}\\
 &=\mathop {\lim }\limits_{x \to 0}1 + \frac{1}{{\frac{1}{2}\ln\frac{{\sqrt x }}{2}}}=1,
\end{align}
where step (a) follows by the approximation $\theta  \to 0,\theta {K_1}\left( \theta  \right) \approx 1 + \frac{{{\theta ^2}}}{2}\left( {\ln\frac{\theta }{2} + {c_0}} \right)$, $c_0=- \frac{{\varphi \left( 1 \right) + \varphi \left( 2 \right)}}{2}$ and $\varphi \left( \cdot \right)$ is the psi function \cite{ye11}.}

\section{Simulations}
In this section, we validate the outage performance of the proposed schemes and the derived system outage probability via $1 \times 10^{6}$ Monte-Carlo simulations.
{\color{black}Unless otherwise specified, the simulation parameters are set as follows. We assume that $\alpha=2.7$ and $\beta=\frac{1}{3}$. Suppose that the carrier frequency used is $915$ MHz and the reference distance $d_0$ is $1$ m \cite{8438944}. Then $\lambda$ can be calculated as $0.33$ m.
The source terminal antenna gain and the relay antenna gain are set as $8$ dBi (https://www.powercastco.com/products/powercaster-transmitter/). According to \cite{7831382}, we assume that $d_A=5$m, ${d_{B}} = 15$m, and $P=30$ dBm. We consider noise variance $\sigma^{2}  = -90$ dBm. The transmission rate is assumed as $U=2\;{\rm{bit/s/Hz}}$ and $\gamma_{\rm{th}}=2^{U}-1$. The energy conversion efficiency is set to be $\eta=0.6$.}

\begin{figure}
  \centering
  \includegraphics[width=0.4\textwidth]{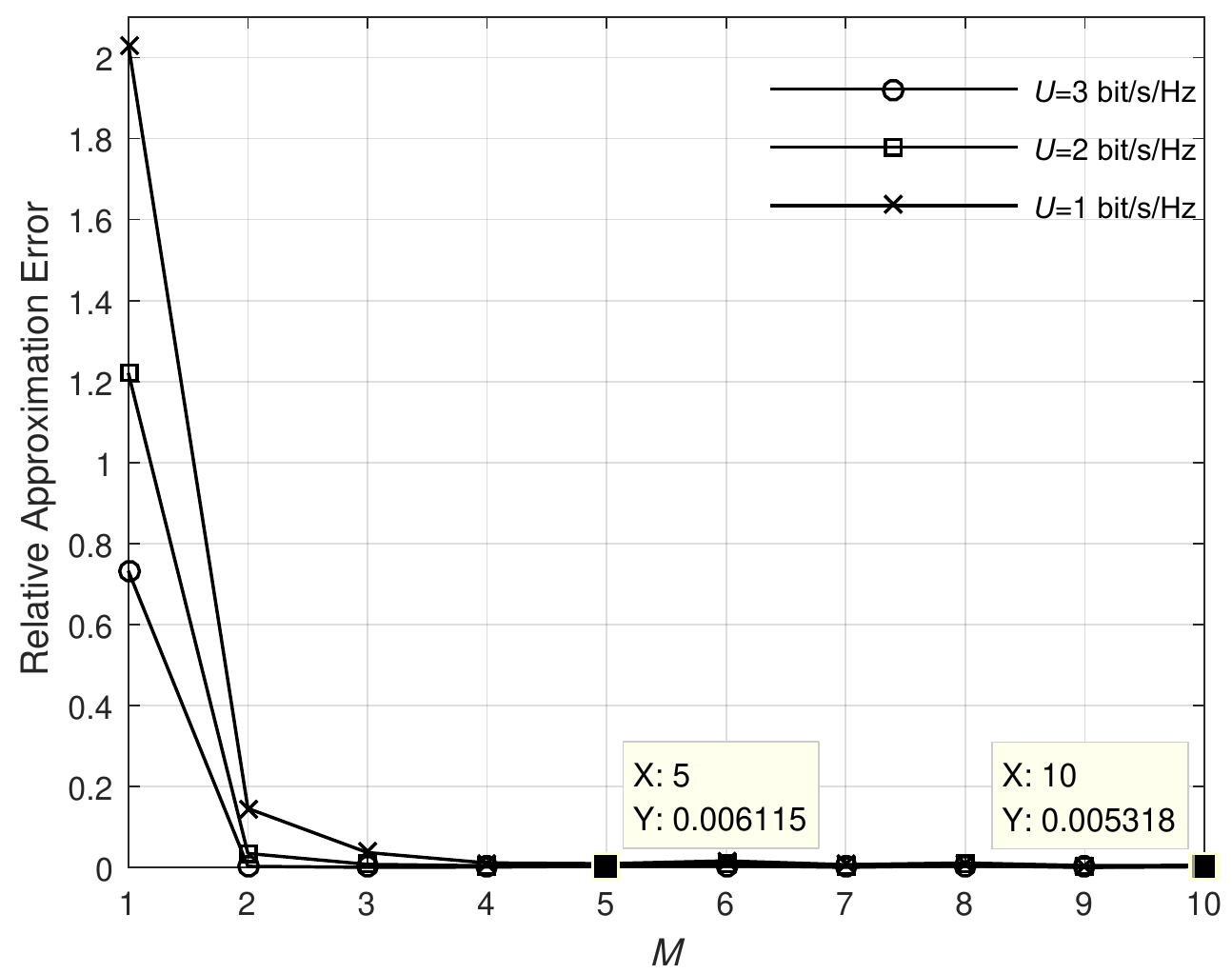}\\
  \caption{Relative approximate error versus parameter $M$.}\label{fig3}
\end{figure}

Fig. \ref{fig3} plots the relative approximate error versus parameter $M$ with different settings of $U$
to illustrate the performance of the Gaussian-Chebyshev quadrature approximation approach. Specifically, according to \cite{6828809}, the relative approximation error can be computed as
\begin{align}\label{111}
\delta  = \left| {\frac{{{\rm{analytical\; result}} - {\rm{simulation \; result}}}}{{{\rm{simulation \;result}}}}} \right|,
\end{align}
where the analytical result is achieved by \eqref{G3} and the simulation result is obtained from Monte-Carlo simulations. As  expected, with the increase of $M$, the relative approximation error approaches zero. For example, when $M=5$, the relative approximation error $\delta$ is $0.006115$, which provides enough accuracy for the system outage probability. Thus, our derived expressions based on the Gaussian-Chebyshev quadrature approximation approach can evaluate the outage performance of the investigated network effectively.

\begin{figure}
  \centering
  \includegraphics[width=0.4\textwidth]{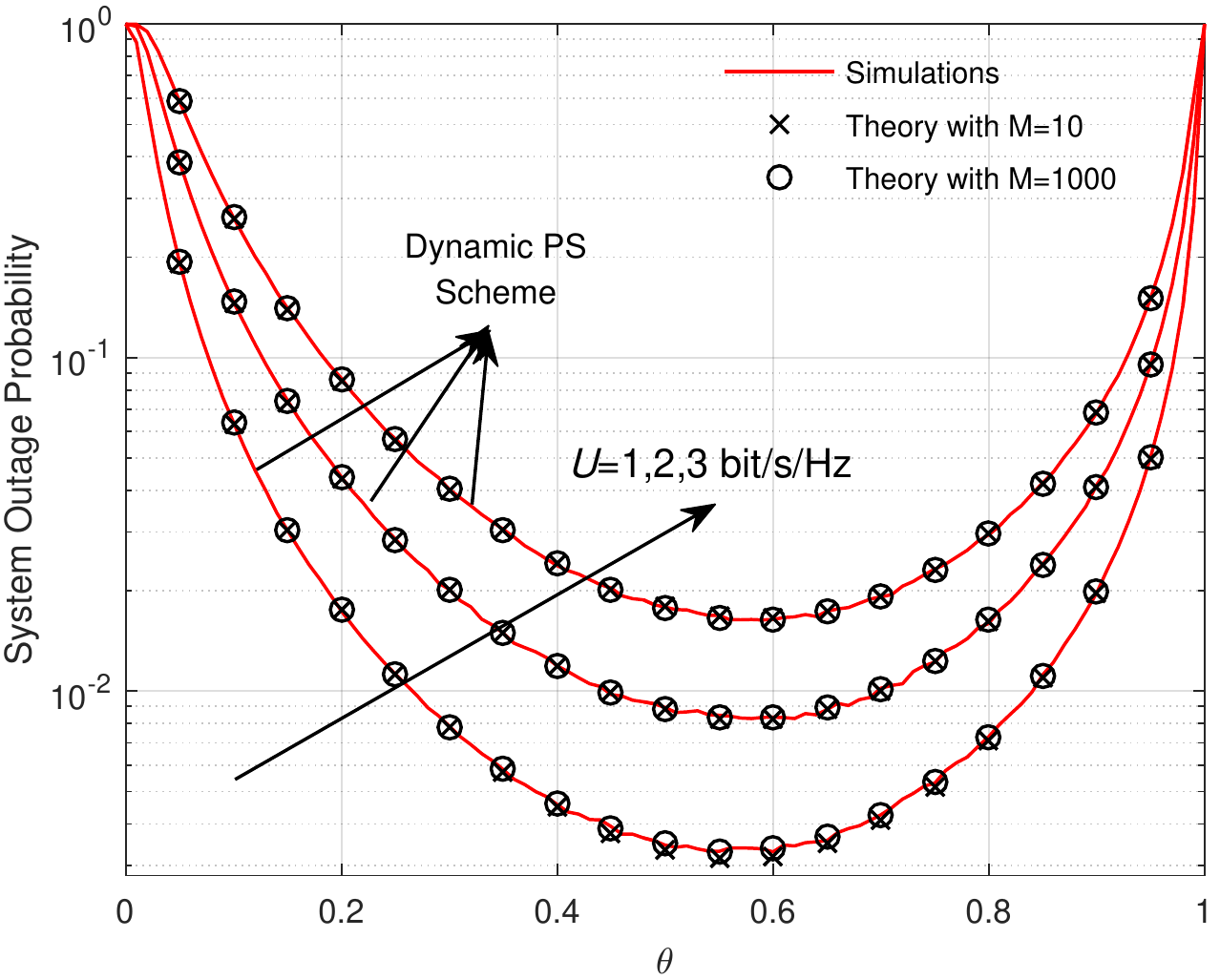}\\
  \caption{{\color{black}System outage probability versus $\theta$ under dynamic PS scheme.}}\label{fig4}
\end{figure}
\begin{figure}
  \centering
  \includegraphics[width=0.4\textwidth]{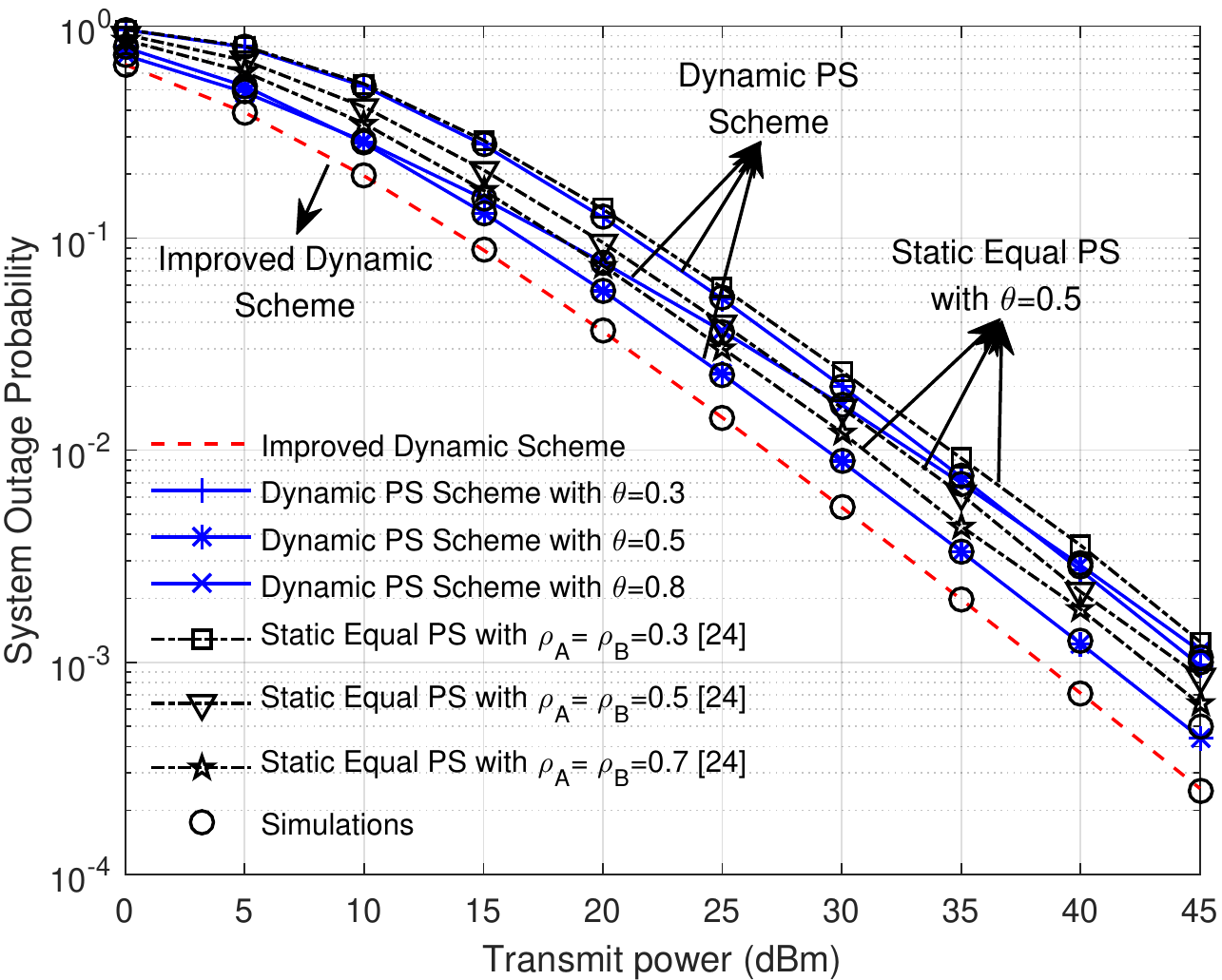}\\
  \caption{{\color{black}System outage probability versus transmit power $P$.}}\label{fig5}
\end{figure}


Fig. \ref{fig4} plots the system outage probability under the dynamic PS scheme versus the power allocation ratio $\theta$ with $U=1,2$, and $3$ bit/s/Hz, respectively.
It can be observed that the theoretical results match with Monte Carlo simulation results with $M=10$ well, which demonstrates the correctness of our derived system outage probability $P_{\rm{out}}^{s}$ in \eqref{G3}.
{\color{black}Another observation is that the system outage probability decreases first, reaches the minimum, and then increases. So there exists an optimal $\theta$, which can bring a minimum $P_{\rm{out}}^{s}$.}
It is worth emphasizing that $\theta=0.5$ assumed in \cite{2017CL} can not yield the minimum $P_{\rm{out}}^{s}$,
and that the minimum $P_{\rm{out}}^{s}$ can be achieved by choosing a proper value of $\theta$. Motivated by this, taking the combining strategy of the relay into account, we propose the improved dynamic scheme to reduce the system outage probability further.

{\color{black}Fig. \ref{fig5} plots the system outage probability as a function of the transmit power, where three schemes are employed,  namely, the proposed improved dynamic scheme, the proposed dynamic PS scheme, and the existing static equal scheme in \cite{2017CL}.}
Specifically, for the improved dynamic scheme, the system outage probability is given by \eqref{s17}. For the dynamic PS scheme, the power allocation ratio is set to be $0.3$, $0.5$, and $0.8$, respectively. For the static equal scheme,
according to \cite{2017CL}, the power allocation ratio is assumed as $0.5$ and $\rho_A=\rho_B$ is set as $0.3$, $0.5$, and $0.7$, respectively. As shown in Fig. \ref{fig5}, it can be observed that the system outage probability under the three schemes decreases with the increase of the transmit power and our derived system outage probability under the improved dynamic scheme also perfectly matches the simulation result, which demonstrates the correctness of \eqref{s17}. With the set of $\theta=0.5$, the dynamic PS scheme enjoys a lower outage probability compared with the static equal scheme in \cite{2017CL}. This is because that the dynamic PS scheme can adjust the PS ratios according to the instantaneous CSI to achieve a lower outage probability. For the dynamic PS scheme with $\theta\neq 0.5$, we can see that the static equal scheme in \cite{2017CL} may be superior to the dynamic PS scheme in terms of outage performance. For example, the static equal scheme with $\rho_A=\rho_B=0.7$ achieves a lower outage probability than the dynamic PS scheme with $\theta=0.3$ or $0.8$. This demonstrates the importance of choosing a proper power allocation ratio $\theta$.
Another observation is that the proposed improved dynamic scheme can achieve the best outage performance among the three schemes.
This is due to the fact that the proposed improved dynamic scheme takes the optimal dynamic PS ratios and the optimal combining strategy into account and can utilize the instantaneous CSI more effectively. {\color{black} Furthermore, it can also be seen that the slope of system outage probability with the dynamic PS scheme (or the improved dynamic scheme) increases with the transmit power and approaches one when the transmit power is large enough, which verifies our diversity gain analysis in Section III.D and Section IV.D.}

\begin{figure}
  \centering
  \includegraphics[width=0.4\textwidth]{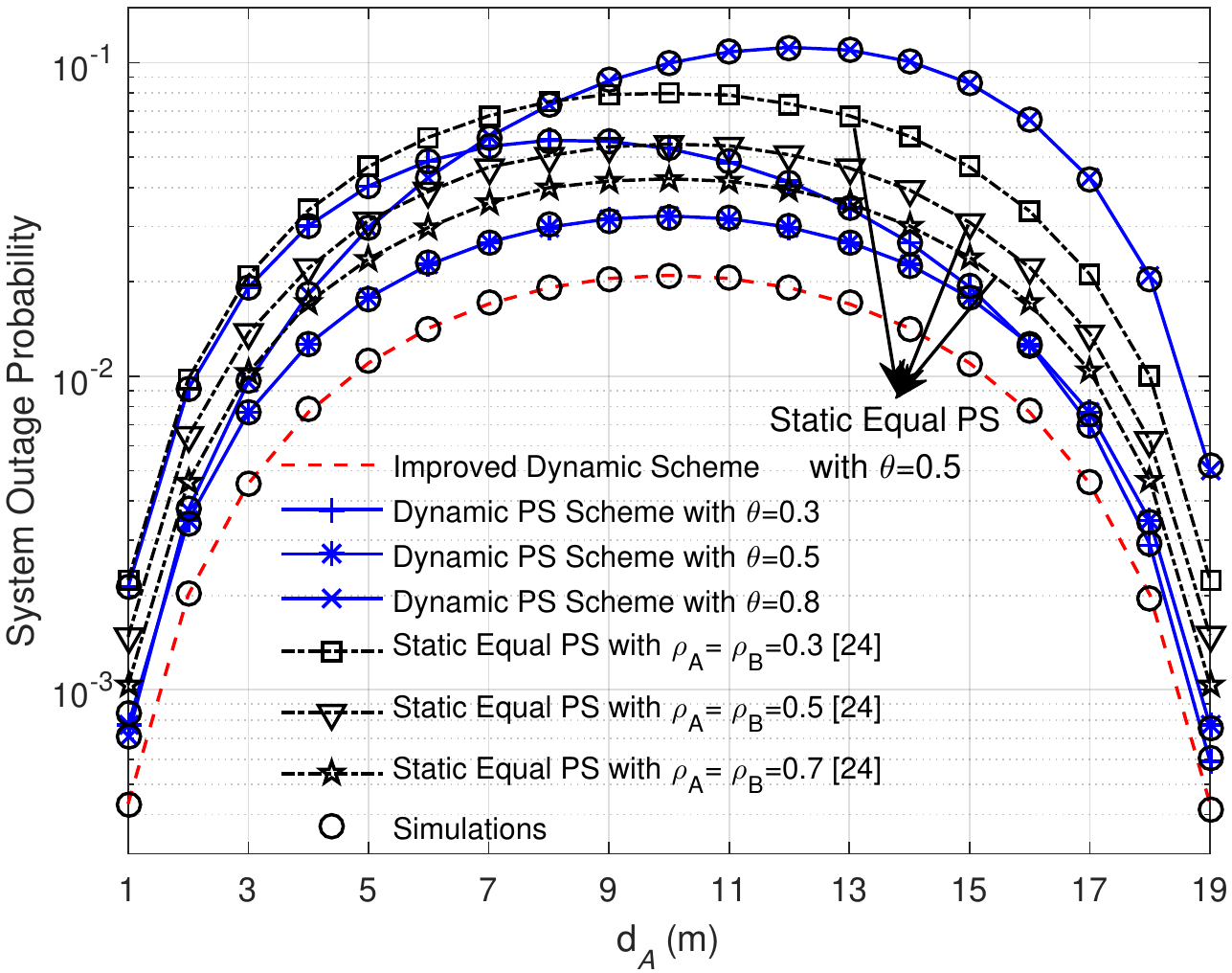}\\
  \caption{{\color{black}System outage probability versus the $A$-$R$ link distance $d_A$ with $U=3\; \rm{bit/s/Hz}$.}}\label{fig6}
\end{figure}

{\color{black}Fig. \ref{fig6} shows the system outage probability for the three schemes versus the $A$-$R$ link distance $d_A$ to depict the effect of the relay location on the system outage probability. The transmission rate $U$ is set as $3\; \rm{bit/s/Hz}$.
As shown in Fig. \ref{fig6}, it can be observed that with the increase of $d_A$, the system outage probability increases first, reaches the maximum value and then decreases. This is due to the fact that the total harvested energy is higher when the relay is closer to either of the terminals. This illustrates that the optimal relay location should be close to either of the terminals to achieve a lower system outage probability.
Another observation is that with the same set of $\theta$, the dynamic PS scheme is always superior to the existing static equal scheme in \cite{2017CL} and the improved dynamic scheme outperforms the dynamic PS scheme in terms of the outage performance.}

\begin{figure}
  \centering
  \includegraphics[width=0.4\textwidth]{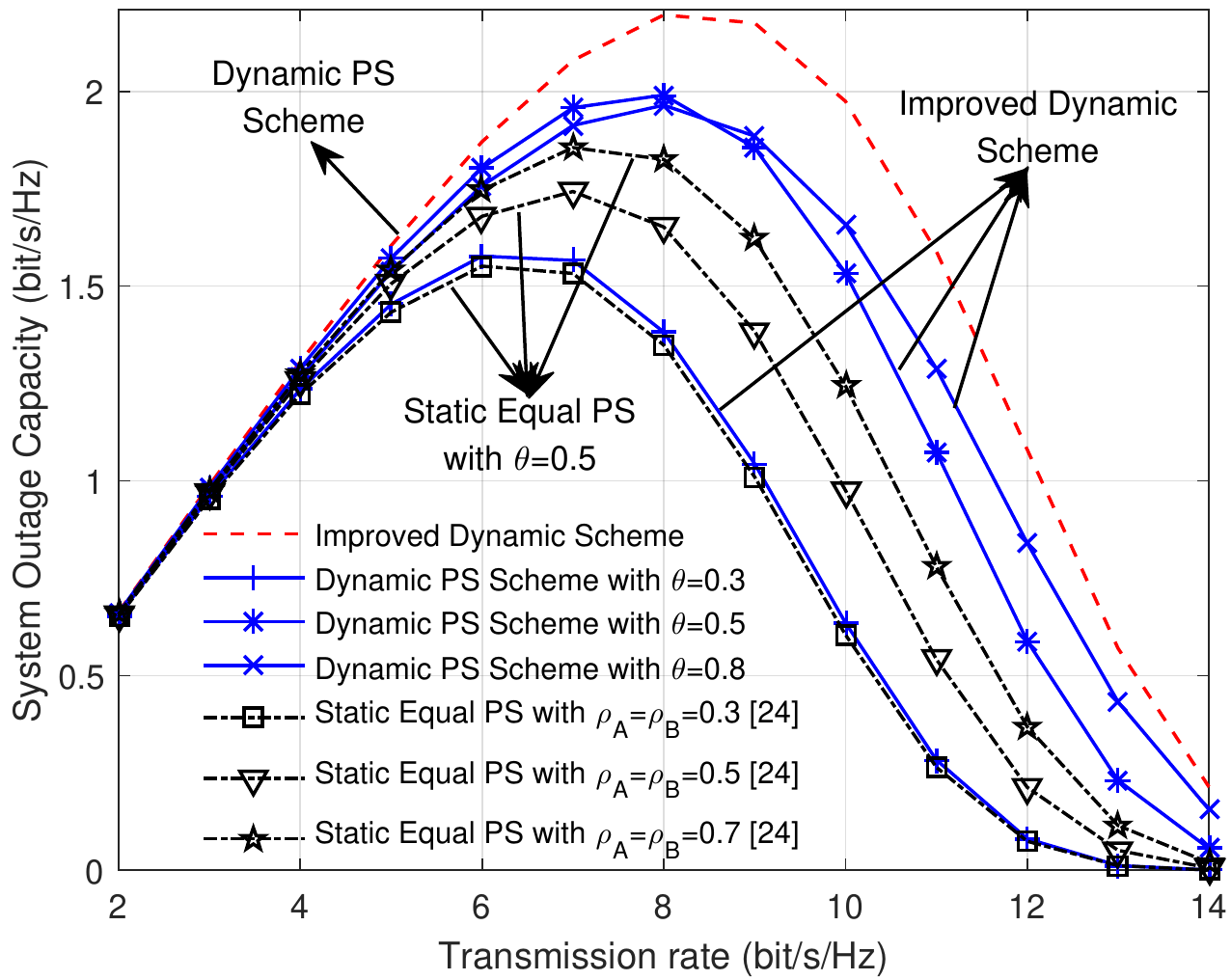}\\
  \caption{{\color{black}System outage capacity versus transmission rate $U$.}}\label{fig7}
\end{figure}

Fig. \ref{fig7} plots the system outage capacity under three schemes versus the transmission rate $U$. {\color{black} Note that, for the improved dynamic scheme, the system outage capacity is given by \eqref{G111}, while for the dynamic PS scheme with $\theta=0.3$, $0.5$ or $0.8$, the system outage capacity is calculated by \eqref{G3} and \eqref{G11}.} It can be seen that with the increase of $U$, the overall system outage capacity increases first, reaches the peak value and then decreases. The reasons are as follows. With a relatively low transmission rate, the transmission rate $U$ is the dominant factor to the outage capacity. Thus, the outage capacity increases with the increasing of $U$. With a larger transmission rate, the receivers may fail to correctly decode the amount of data. In this case, the outage probability becomes the dominant factor and the outage capacity decreases with the increasing outage probability. Besides, we can see that a well-designed $U$  can bring a higher outage capacity.
Another observation is that the improved dynamic scheme can provide a significant performance gain over the dynamic PS schemes and the static equal PS schemes, while with the same set of $\theta$, the dynamic PS scheme always outperforms the static equal PS scheme.

\begin{figure}
  \centering
  \includegraphics[width=0.4\textwidth]{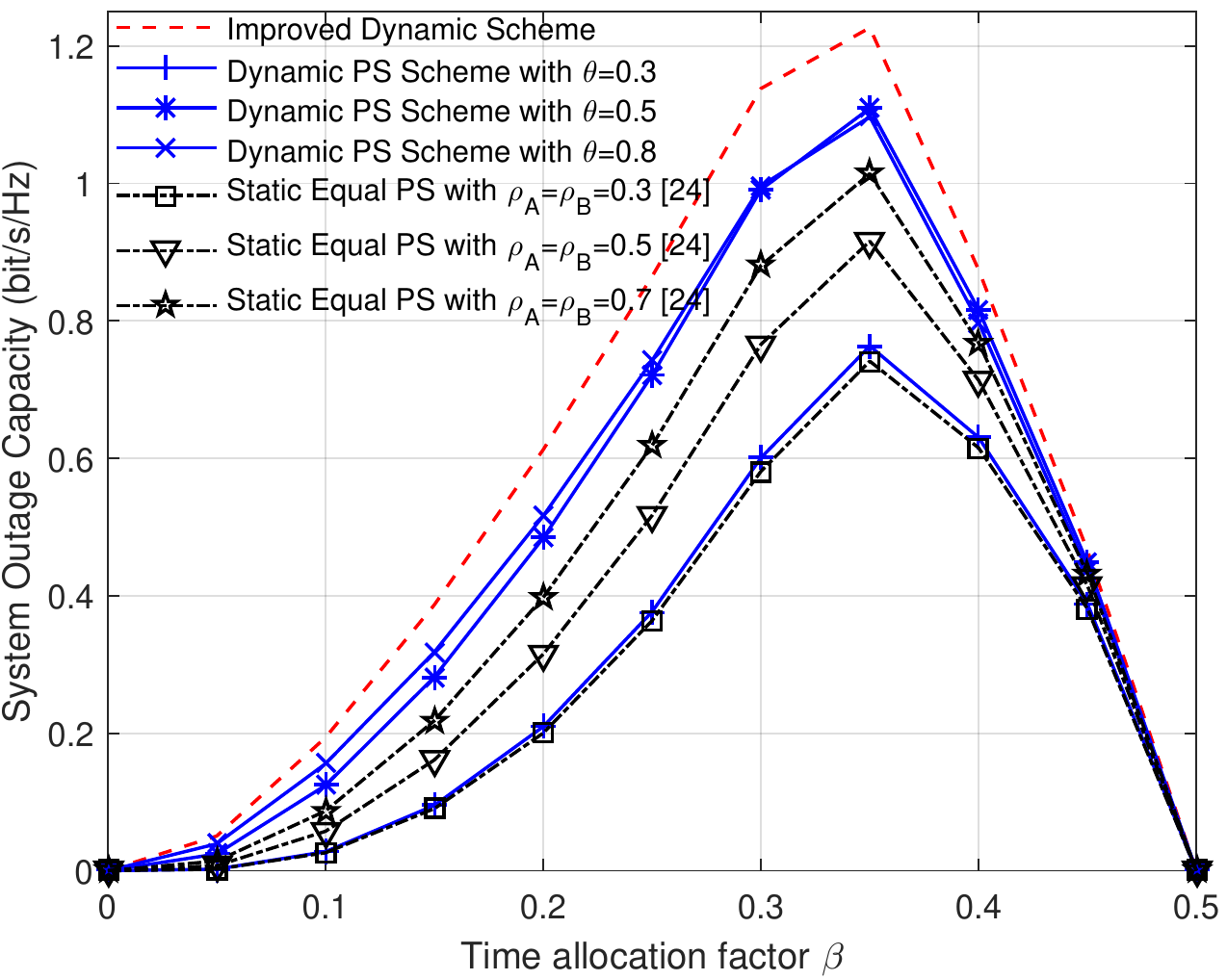}\\
  \caption{{\color{black}System outage capacity versus time allocation ratio $\beta$ with $P=20$ dBm and $U=5\; \rm{bit/s/Hz}$.}}\label{fig8}
\end{figure}

\begin{figure}
  \centering
  \includegraphics[width=0.4\textwidth]{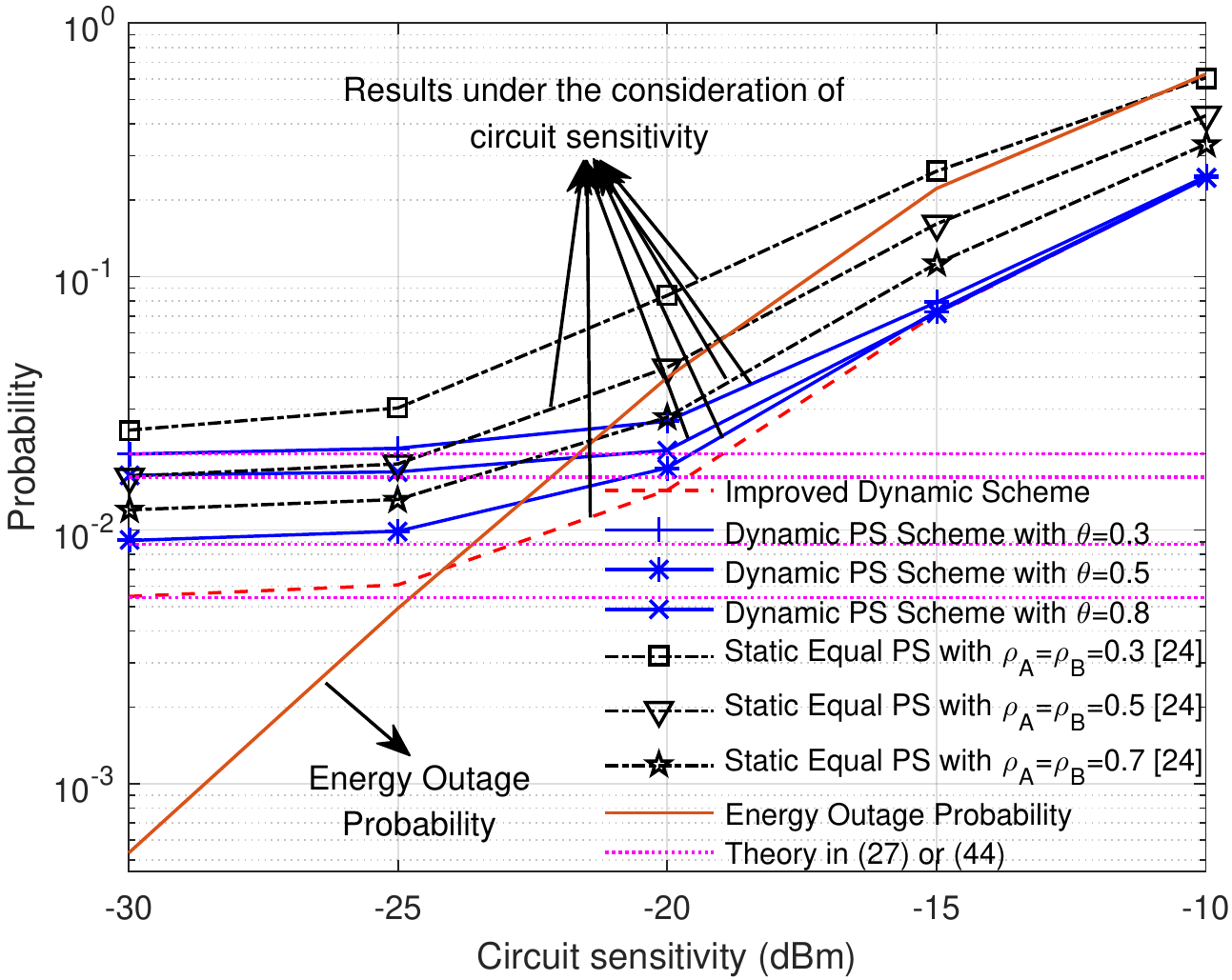}\\
  \caption{{\color{black}System outage probability and energy outage probability versus circuit sensitivity $P_{\rm{th}}$.}}\label{fig9}
  \vspace*{-15pt}
\end{figure}

Fig. \ref{fig8} illustrates the relationship between the system outage capacity and the time allocation ratio $\beta$ with above three schemes considered. {\color{black}We set $P=20$ dBm and $U=5\; \rm{bit/s/Hz}$.} One observation is that the system outage capacity increases with the increase of $\beta$ and then decreases. There exists an optimal time allocation ratio $\beta^*$ for each scheme to achieve the maximum system outage capacity and $\beta^*\geq \frac{1}{3}$. The reason is as follows. When $0<\beta\leq \frac{1}{3}$, the effective transmission time for the system outage capacity is given by $\beta T$ and the system outage capacity increases with increase of $\beta$ due to the fact that a larger $\beta$  yields larger $\gamma_{RA}$ and $\gamma_{RB}$, leading to a lower outage probability. When $\beta>\frac{1}{3}$, the effective transmission time is given by $(1-2\beta) T$. For a larger $\beta$, the effective transmission time becomes the dominant factor to the capacity. Thus, the system outage capacity shows a downward trend. On the other hand, we can also see that the improved dynamic scheme can achieve the highest capacity among the three schemes and with the same set of $\theta$, the dynamic PS scheme is superior to the static equal PS scheme.

{\color{black}Fig. \ref{fig9} plots the energy outage probability and the system outage probability versus the circuit sensitivity $P_{\rm{th}}$. According to \cite{6951347}, the range of $P_{\rm{th}}$ is set to be $[-30, -10]$ dBm. The energy outage probability is defined as the probability that harvested energy at the relay is $0$. We use $P_{\rm{eo}}$ to denote the energy outage probability. Then we have $P_{\rm{eo}}=\mathbb{P}({P{\rho _A}|{h_A}{|^2}{\Lambda _A}d_A^{ - \alpha } < {P_{{\rm{th}}}}},{P{\rho _B}|{h_B}{|^2}{\Lambda _B}d_B^{ - \alpha } < {P_{{\rm{th}}}}})=\left[ {1 - \exp \left( { - {a_A}\left( {\frac{{{P_{\rm{th}}}}}{P} + \varpi } \right)} \right)} \right]\left[ {1 - \exp \left( { - {a_B}\left( {\frac{{{P_{\rm{th}}}}}{P} + \varpi } \right)} \right)} \right]$. For the system outage probability, two cases are considered: (i) the case with $P_{\rm{th}}\neq 0$, (ii) the case with $P_{\rm{th}}= 0$.
It can be observed that both $P_{\rm{eo}}$ and the system outage probability increase with the increase of $P_{\rm{th}}$ and our derived expressions, \eqref{G3} and \eqref{s17}, provide lower bounds for the system outage probability in a practical scenario with the circuit sensitivity considered.
Specifically, when $P_{\rm{th}}$ is small, the results obtained by our derived expressions of system outage probability are very close to those obtained by considering $P_{\rm{th}}\neq 0$. Besides, we can also see that considering the circuit sensitivity,  with the same set of $\theta$, the dynamic PS scheme is still superior to the existing static equal scheme in \cite{2017CL} and the improved dynamic scheme still outperforms the dynamic PS scheme in terms of the outage performance.}

\section{Conclusions}
In this paper, we have proposed two schemes: dynamic PS scheme and improved dynamic scheme, to minimize the system outage probability for the SWIPT enabled three-step DF TWRNs. Specifically, for each scheme, we have derived the optimal solutions in closed forms and further derived the analytical expressions for the optimal system outage probability and capacity.
Simulation results validate the correctness of  our derived outage probabilities and capacities. The impacts of various  parameter settings, e.g., the relay location, the time allocation ratio and  the transmission rate, on the system outage performance of  SWIPT enabled three-step DF TWRNs have been studied.  Several insights have been obtained. First, the proposed schemes are superior to the existing static scheme and the improved dynamic scheme enjoys a lower system outage probability than the dynamic PS scheme. Second, a considerable performance gain can be obtained after carefully selecting a proper time allocation ratio and transmission rate.
\section*{Appendix}
\subsection{Derivation of $F_{t_2}(t)$}
According to the definition of $F_{t_2}(t)$, we have
\begin{align}\notag\label{c1}
&F_{t_2}(t)=\mathbb{P}(t_2\leq t)=\mathbb{P}\left[ {x \le \left( {t - yZ_B^{ - 1 }} \right)Z_A ,y \le tZ_B } \right]\\ \notag
&=\int_0^{tZ_B } {\left[ {1 - \exp \left( { - {a_A}\left( {t - yZ_B^{ - 1 }} \right)} \right)} \right]} \frac{{\exp \left( { - {y \mathord{\left/
 {\vphantom {y {{\lambda _B}}}} \right.
 \kern-\nulldelimiterspace} {{\lambda _B}}}} \right)}}{{{\lambda _B}}}dy\\
 &={1 - {e^{ - {a_B}t}} - \frac{{{e^{ - {a_A}t}}}}{{{\lambda _B}}}\int_0^{tZ_B } {\exp \left( {\frac{{{a_A} - {a_B}}}{{{a_B}{\lambda _B}}}y} \right)} dy,}\!\!
\end{align}
where $x=|{h_A}{|^2}$, $y=|{h_B}{|^2}$ and ${a_i} = \frac{{Z_i }}{{{\lambda _i}}}$.
When ${a_A}={a_B}$, \eqref{c1} can be computed as 
\begin{align}
F_{t_2}(t)={1 - {e^{ - {a_B}t}} - {a_B}t{e^{ - {a_A}t}}}.
\end{align}
For the case with ${a_A}\neq{a_B}$, \eqref{c1} is given by
\begin{align}
F_{t_2}(t)={1 - {e^{ - {a_B}t}} - \frac{{{a_B}}}{{{a_A} - {a_B}}}\left( {{e^{ - {a_B}t}} - {e^{ - {a_A}t}}} \right)}.
\end{align}
Thus, $F_{t_2}(t)$ can be rewritten as (38).

\subsection{Derivation of $F_{t_3}(t)$}
Similarly, $F_{t_3}(t)$ is given by
\begin{align}\notag
F_{t_3}(t)&=\mathbb{P}(x\geq\frac{1}{yt})=\frac{1}{{{\lambda _B}}}\int_0^{ + \infty } {\exp \left( { - \frac{1}{{{\lambda _A}ty}} - \frac{y}{{{\lambda _B}}}} \right)dy}\\
&
=\frac{1}{{{\lambda _B}}}\sqrt {\frac{{4{\lambda _B}}}{{{\lambda _A}t}}} {K_1}\left( {\sqrt {\frac{4}{{{\lambda _A}{\lambda _B}t}}} } \right),
\end{align}
where ${K_1}\left(  \cdot  \right)$ is the modified Bessel function of the second kind.
The proof is completed.
\ifCLASSOPTIONcaptionsoff
  \newpage
\fi
\bibliographystyle{IEEEtran}
\bibliography{refa}

\begin{IEEEbiography}[{\includegraphics[width=1in,height=1.25in,clip,keepaspectratio]{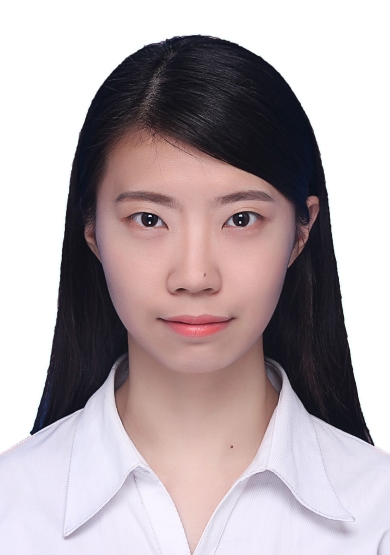}}]{Liqin Shi}
received the B.S. degree in electronic information  engineering from Sichuan University, Chengdu, China, in 2015. She is currently working toward the Ph.D. degree with the State Key Laboratory of Integrated Service Networks, Xidian University, and a Joint Ph.D. Student with the Department of Electrical and Computer Engineering, Utah State University, UT, USA, under the supervision of Professor Rose Qingyang Hu. Her research interests include wireless energy harvesting, D2D communications, etc. She has published multiple papers in IEEE Transactions on Vehicular Technology, IEEE ICC, etc.
\end{IEEEbiography}

\begin{IEEEbiography}[{\includegraphics[width=1in,height=1.25in,clip,keepaspectratio]{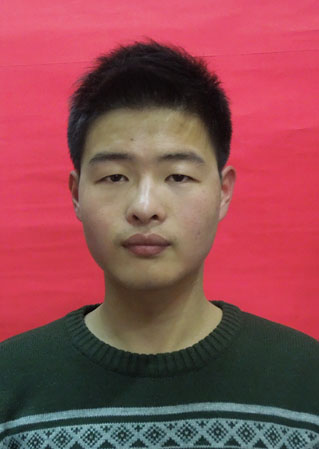}}]{Yinghui Ye}
 received the M.S. degree in communication and information system from Xi'an University
of Posts and Telecommunications, Xi¡¯an, China, in 2016. Currently, he is working toward the Ph.D.
degree with the State Key Laboratory of Integrated Service Networks, Xidian University, and a Joint
Ph.D. Student with the Department of Electrical and Computer Engineering, Utah State University,
UT, USA, under the supervision of Professor Rose Qingyang Hu. His research interests include cognitive
radio networks, relying networks, and wireless energy harvesting. He has authored/coauthored more than twenty technical articles in international journals and proceedings. He is also a reviewer of multiple
international Journals and conferences, including the IEEE Transactions on Wireless Communications, the IEEE Transactions on Communications, the IEEE Transactions on Vehicular Technology, the IEEE Internet of Things Journal, the IEEE Communications Letters, the IEEE VTC, etc.
 \end{IEEEbiography}

\begin{IEEEbiography}[{\includegraphics[width=1in,height=1.25in,clip,keepaspectratio]{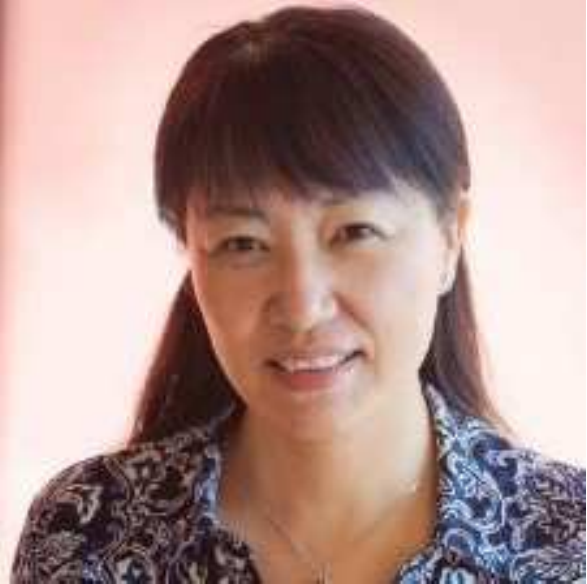}}]{RoseQingyang Hu} is a Professor of Electrical and Computer Engineering Department at Utah State University. She received her B.S. degree from University of Science and Technology of China, her M.S. degree from New York University, and her Ph.D. degree from the University of Kansas. She has more than 10 years of R\&D experience with Nortel, Blackberry and Intel as a technical manager, a senior wireless system architect, and a senior research scientist, actively participating in industrial 3G/4G technology development, standardization, system level simulation and performance evaluation. Her current research interests include next-generation wireless communications, wireless system design and optimization, green radios, Internet of Things, Cloud computing/fog computing, multimedia QoS/QoE, wireless system modeling and performance analysis. She has published over 180 papers in top IEEE journals and conferences and holds numerous patents in her research areas. Prof. Hu is an IEEE Communications Society Distinguished Lecturer Class 2015-2018 and the recipient of Best Paper Awards from IEEE Globecom 2012, IEEE ICC 2015, IEEE VTC Spring 2016, and IEEE ICC 2016.
 \end{IEEEbiography}

 \begin{IEEEbiography}[{\includegraphics[width=1in,height=1.25in,clip,keepaspectratio]{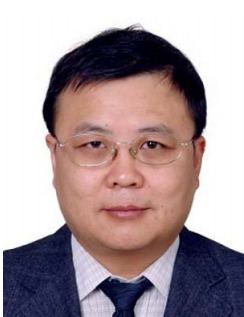}}]{Hailin Zhang } (M'98) received the B.S. and M.S. degrees from Northwestern Polytechnic University, Xi'an, China, in 1985 and 1988, respectively, and the Ph.D. degree from Xidian University, Xi'an, China, in 1991. In 1991, he joined School of Telecommunications Engineering, Xidian University, where he is now a Senior Professor. He is also currently the Director of Key Laboratory in Wireless Communications sponsored by China Ministry of Information Technology, a key member of State Key Laboratory of Integrated Services Networks, one of the state government specially compensated scientists and engineers, a field leader in Telecommunications and Information Systems in Xidian University, an Associate Director for National 111 Project. Prof. Zhang's current research interests include key transmission technologies and standards on broadband wireless communications for 5G wireless access systems. He has published more than 100 papers in journals and conferences.
\end{IEEEbiography}

\end{document}